\documentclass[11pt]{article}
\usepackage[a4paper,margin=28mm]{geometry}
\usepackage[T1]{fontenc}
\usepackage{lmodern,microtype,amsmath,amssymb,amsthm,mathtools,booktabs,longtable,array,enumitem}
\usepackage[dvipsnames]{xcolor}
\usepackage[colorlinks=true,linkcolor=MidnightBlue,citecolor=MidnightBlue,urlcolor=MidnightBlue]{hyperref}
\hypersetup{pdftitle={The determinantal complexity of the 4x4 permanent is 15},pdfauthor={Karthik Sheshadri},pdfkeywords={permanent, determinant, determinantal complexity}}
\usepackage{xurl}
\newtheorem{theorem}{Theorem}[section]
\newtheorem{proposition}[theorem]{Proposition}
\newtheorem{lemma}[theorem]{Lemma}

\theoremstyle{definition}
\theoremstyle{remark}\newtheorem{remark}[theorem]{Remark}
\newcommand{\C}{\mathbb C}\newcommand{\Q}{\mathbb Q}\newcommand{\PP}{\mathbb P}
\newcommand{\per}{\operatorname{per}}\newcommand{\dc}{\operatorname{dc}}
\newcommand{\rk}{\operatorname{rank}}\newcommand{\im}{\operatorname{im}}
\newcommand{\Sing}{\operatorname{Sing}}\newcommand{\tr}{\operatorname{tr}}
\newcommand{\diag}{\operatorname{diag}}\newcommand{\Gr}{\operatorname{Gr}}
\newcommand{\Span}{\operatorname{span}}\newcommand{\Sym}{\operatorname{Sym}}
\newcommand{\Pf}{\operatorname{Pf}}\newcommand{\adj}{\operatorname{adj}}
\newcommand{\End}{\operatorname{End}}
\newcommand{\file}[1]{\nolinkurl{#1}}
\allowdisplaybreaks[2]
\setlist{nosep}
\title{The determinantal complexity of the $4\times4$ permanent is $15$}
\author{Karthik Sheshadri}
\date{}
\begin{document}
\maketitle
\begin{abstract}
The best known lower bound on the determinantal complexity of the $4\times4$ permanent is 10. In this paper, I prove that this value is exactly 15: the $4\times4$ permanent is the determinant of a $15\times15$ matrix of affine linear forms, and cannot be that of any smaller matrix. To do so, I must establish both a lower and upper bound. To establish the upper bound, I use Grenet's construction of size $2^n-1$. Hence that construction is optimal at $n=4$ (and $n=3$). This now constitutes the first known exact value of the determinantal complexity of the permanent at $n=4$, improving over prior knowledge in which the exact value was known only for $n=3$. Recall that the general quadratic bound only gives 8 for $n=4$.

To establish the lower bound, I work from a normal form at a zero of the permanent. In the first row (and column) of the normal form, the linear forms vanish on a linear space inside the singular locus of the permanent. A hessian inequality bounds the size of the matrix in terms of that space. Then, a matrix of size 14 forces the space into any of nine coordinate types. The forcing is either up to symmetry, or else into a family that degenerates to one of them. I then rule out every case. All involved computations are exact.

In respect of the claim, I have made all certificates public. This version of the paper gives the full proof of every step.
\end{abstract}
\begingroup
\small
\tableofcontents
\endgroup

\section{Introduction}
The determinantal complexity $\dc(f)$ of a polynomial $f$ is the smallest $m$ for which $f=\det A$, where $A$ is an $m\times m$ matrix whose entries are affine linear forms in the variables of $f$. Valiant showed that every polynomial computed by a small formula has small determinantal complexity, and that the permanent is complete for the class VNP~\cite{Valiant}. His conjecture that $\dc(\per_n)$ grows faster than any polynomial in $n$ is the determinant versus permanent form of the VP versus VNP question.

I consider the exact computation of small values of the permanent as revealing of its behavior and of the pattern of growth of the permanent for larger, and asymptotically larger, values of $n$. It is my hope that a close study of the techniques that yield these values may one day cast a long enough shadow that at the least, a rough shape of the mathematics behind it may be deduced.

The known bounds are far apart. Mignon and Ressayre proved $\dc(\per_n)\geq n^2/2$ over $\C$ by comparing Hessians~\cite{MR}, and Cai, Chen and Li extended this to every characteristic other than $2$~\cite{CCL}. For $n=3$, Alper, Bogart and Velasco proved that $\dc(\per_3)\geq7$~\cite{ABV}. On the upper bound side, Grenet gave a representation of size $2^n-1$~\cite{Grenet}. These coincide at $n=3$. Landsberg and Ressayre showed that $2^n-1$ is optimal among representations that respect a large part of the symmetry group of the permanent~\cite{LR}, which leaves open whether a representation without that symmetry can be smaller. At $n=4$ the quadratic bound gives $8$ against Grenet's $15$. An earlier stage of the present author's own analysis had recorded a best lower bound for $n=4$ as $10$.

To the best of my knowledge, this is the first article that settles the exact case of $n=4$ from both directions.

\begin{theorem}\label{thm:main}
$\dc(\per_4)=15$ over $\C$.
\end{theorem}

The proof has four parts. Firstly, the singular locus of the permanent hypersurface has dimension $8$. This forces every representation to have corank exactly one at every zero of $\per_4$. It gives a normal form at the origin in which $\per_4=bD^2c$, for a row $b$, a column $c$ and a square matrix $D$ of linear forms.

Secondly, the forms in $b$ and $c$ vanish together on a linear space $K$, which I call the endpoint kernel. The endpoint kernel lies in the singular locus. Let $\rho$ be the rank of the Hessian of $\per_4$ at a generic point of $K$. A short argument with a quadratic form then shows that the size of the matrix is at least $17-\dim K+\lceil\rho/2\rceil$.

Thirdly, at size $14$ this inequality leaves nine coordinate kernels up to symmetry. The torus acting on rows and columns degenerates every other surviving kernel to one of them. A local computation near these nine then shows that the forcing is either up to symmetry, or else into a family that degenerates to one of them. The families consist of the hyperplanes of a row, of a $2\times3$ block, and of two rows.

Finally, I exclude each surviving kernel at size $14$. The hardest case is the kernel spanned by a single row. There the Hessian vanishes, and the inequality gives nothing beyond $13$. Table~\ref{tab:mechanisms} summarizes how each kernel is excluded.

\begin{table}[ht]
\centering\small
\begin{tabular}{p{.12\textwidth}p{.27\textwidth}p{.41\textwidth}l}\toprule
Container & Kernels & Mechanism & Section\\\midrule
one row & row less a cell, and every hyperplane of a row & stabilizer of the cubic $F_z$; nilpotence of $D$ on the kernel; irreducibility of $F_z$ & \ref{sec:rowthree}\\
two rows & full $2\times3$ block and its hyperplanes & no linear syzygies among $2\times2$ subpermanents & \ref{sec:block2x3}, \ref{sec:block2x3hyp}\\
cross & cross with arms of two & rank of the quotient map against the Hessian radical & \ref{sec:arms2}\\
two rows & row and two cells & rank of the specialized Hessian block & \ref{sec:row2cells}\\
cross & $K_6$ & equality case of the inequality; irreducibility of $\per_3$ & \ref{sec:K6}\\
two rows & two rows, two rows less a cell, and every hyperplane of two rows & five channel factorization of $B$; Clifford reduction; odd factor obstruction & \ref{sec:tworows}\\
one row & one row & invariant subspaces of a nilpotent pencil & \ref{sec:fullrow}\\\bottomrule
\end{tabular}
\caption{How each admissible kernel is excluded at size $14$. Kernels inside a container include the noncoordinate ones.}
\label{tab:mechanisms}
\end{table}

The argument depends on the characteristic. In characteristic $2$ the permanent equals the determinant and $\dc(\per_4)=4$. The step that fails there is the first one, because the singular locus of $\det_4$ has dimension $12$.

The first version of this paper, which also appeared as ECCC report TR26-309, condensed several of these steps and referred to the certificate archives for them. In this version every step is argued in full. The computer still enters, but only through finite exact calculations: an enumeration of coordinate subspaces, ranks of explicit integer matrices, and polynomial identities. Each of them is specified in the text, and the two largest are printed in the appendices.

The remainder of this paper is organized as follows. Section~\ref{sec:foundations} gives the upper bound and derives the normal form and the endpoint inequality. Section~\ref{sec:classification} classifies the kernels that survive the inequality. Sections~\ref{sec:small}, \ref{sec:tworows} and~\ref{sec:fullrow} exclude them, and Section~\ref{sec:audit} lists the certificates. Appendices~\ref{sec:polynomialappendix} and~\ref{sec:minorappendix} contain the data of the two largest computations.

\section{The upper bound and the endpoint inequality}
\label{sec:foundations}
This section proves the upper bound, reduces any representation of $\per_4$ to a normal form at the origin, and proves the inequality that the rest of the paper uses. We write $X=(x_{ij})_{1\leq i,j\leq4}$ for the $4\times4$ matrix of variables, and put
\[
 f(X)=\per_4(X)=\sum_{\sigma\in S_4}\prod_{i=1}^4x_{i,\sigma(i)}.
\]
A representation of size $m$ is an $m\times m$ matrix $A(X)$ of affine linear forms in the sixteen variables with $\det A=f$. All vector spaces, polynomial rings and varieties are over $\C$. The rank of a polynomial matrix means its rank over the fraction field, which is also its rank on a nonempty Zariski open subset of its parameter space.

\subsection{The upper bound}\label{sec:upper}
The upper bound is Grenet's construction~\cite{Grenet}, which we recall for $n=4$. Index fourteen states by the nonempty proper subsets $S$ of $\{1,2,3,4\}$. Order the states first by their cardinality. There are four states of size one, six of size two, and four of size three. Define a row $b$, a column $c$, and a $14\times14$ matrix $D$ by
\[
 b_S=\begin{cases}x_{1j}&S=\{j\},\\0&\text{otherwise},\end{cases}
 \qquad
 c_S=\begin{cases}x_{4j}&S=\{1,2,3,4\}\setminus\{j\},\\0&\text{otherwise},\end{cases}
\]
and
\[
 D_{S,T}=\begin{cases}
 x_{|S|+1,j}&T=S\cup\{j\},\ j\notin S,\ |S|=1\text{ or }2,\\
 0&\text{otherwise}.
 \end{cases}
\]
Every nonzero entry of $D$ increases subset cardinality by one. Thus $D^3=0$, $I-D$ is upper triangular with diagonal entries one, and
\[
 \det(I-D)=1,\qquad (I-D)^{-1}=I+D+D^2.
\]
The product $bc$ is zero because $b$ is supported on states of size one and $c$ on states of size three. The product $bDc$ is also zero because one step does not connect these two layers. A term of $bD^2c$ corresponds to an ordering $(j_1,j_2,j_3,j_4)$ of the four columns, and its weight is
\[
 x_{1j_1}x_{2j_2}x_{3j_3}x_{4j_4}.
\]
Each ordering occurs exactly once, so $bD^2c=f$. The Schur complement now gives
\[
 \det\begin{pmatrix}0&-b\\c&I-D\end{pmatrix}
 =\det(I-D)b(I-D)^{-1}c=bD^2c=f.
\]
This proves $\dc(f)\leq15$.

\subsection{Polynomials attached to the permanent}\label{sec:targetfacts}
For $u,v\in\C^4$, define the symmetric $4\times4$ matrix $B(u,v)$ by
\[
 B(u,v)_{ii}=0,\qquad B(u,v)_{ij}=u_kv_l+u_lv_k
 \quad(\{i,j,k,l\}=\{1,2,3,4\}).
\]
Then, for four row vectors $a,u,v,c$,
\[
 \per_4(a,u,v,c)=a^TB(u,v)c.
\]
For a single vector $a\in\C^4$, write
\[
 B_1(a)=B(\mathbf1,a),\qquad
 B_1(a)_{ij}=\sum_{k\notin\{i,j\}}a_k\quad(i\ne j).
\]
The subscript distinguishes this linear matrix from the bilinear matrix $B(u,v)$. The trilinear cubic obtained by setting one permanent row to $\mathbf1$ is
\[
 F(a,b,c)=\per_4(\mathbf1,a,b,c)
 =\sum_{i,j,k\text{ pairwise distinct}}a_i b_j c_k
 =b^TB_1(a)c.
\]
For $d=(d_1,d_2,d_3)$ we also use
\begin{equation}\label{eq:Cmatrix}
 C(d)=\begin{pmatrix}0&d_3&d_2\\d_3&0&d_1\\d_2&d_1&0\end{pmatrix},
 \qquad \det C(d)=2d_1d_2d_3.
\end{equation}

\begin{lemma}\label{lem:cubicirreducible}
For every nonzero $z\in\C^4$, the cubic
\[
 F_z(a,b,c)=\per_4(z,a,b,c)
\]
is irreducible. The ordinary $3\times3$ permanent is irreducible as well.
\end{lemma}
\begin{proof}
A polynomial homogeneous of degree one in each of three disjoint groups of variables can factor only into multihomogeneous factors. To see this, fix one group and compare the highest and lowest degrees in that group in a product. In an integral domain the highest degrees add and the lowest degrees add. The width between them therefore adds. A product of width zero has factors of width zero. Apply this argument to each group.

A nontrivial factorization of a polynomial of multidegree $(1,1,1)$ partitions the three row groups between its factors. At least one factor must consequently be a linear form in a single row, independent of the other rows. Such a factor makes all partial derivatives in that row constant multiples of one polynomial.

For $\per_3$, the three derivatives in any row are the three distinct $2\times2$ permanents in the other two rows. They are linearly independent: each contains a monomial not contained in the other two. Hence the required proportionality cannot occur.

For $F_z$, choose $j$ with $z_j\ne0$ and set column $j$ of all three lower rows to zero. Every term using $z_l$ for $l\ne j$ then vanishes, because a lower row would have to use column $j$. The remaining polynomial is $z_j\per_3$ on the other three columns. Consequently the derivatives in each lower row have span of dimension at least three before this specialization. They cannot all be proportional. The multigrading argument therefore excludes every nontrivial factorization.
\end{proof}

\begin{lemma}\label{lem:crossdet}
Let $e_i$ denote elementary symmetric functions in four variables. Then
\[
 \det B_1(a)=-4\bigl(e_1(a)e_3(a)-4e_4(a)\bigr).
\]
This determinant is irreducible. The polynomial $\Delta(u,v)=\det B(u,v)$ is irreducible of total degree eight, and remains irreducible as a polynomial in $v$ over $\C(u)$.
\end{lemma}
\begin{proof}
Expanding the determinant gives the displayed identity. Set
\[
 q(a)=e_1(a)e_3(a)-4e_4(a),\quad
 S=a_2+a_3+a_4,\quad P=a_2a_3+a_2a_4+a_3a_4,\quad R=a_2a_3a_4.
\]
As a quadratic in $a_1$,
\[
 q=P a_1^2+(SP-3R)a_1+SR.
\]
Its coefficients have greatest common divisor one. Indeed a common divisor divides $P$ and $SP-3R$, hence divides both $P$ and $R$, which are relatively prime. The discriminant is
\[
 (SP-3R)^2-4PSR=(SP-R)(SP-9R).
\]
The first factor equals $(a_2+a_3)(a_2+a_4)(a_3+a_4)$. None of these three distinct linear factors divides the second: for example, setting $a_2=-a_3$ in $SP-9R$ gives $8a_3^2a_4$. Thus the discriminant has an irreducible factor of odd multiplicity and is not a square in $\C(a_2,a_3,a_4)$. A quadratic over a field of characteristic zero is reducible precisely when its discriminant is a square. Gauss's lemma now proves the irreducibility of $q$.

On the open set $u_1u_2u_3u_4\ne0$, put $U=\prod_i u_i$ and $z_i=v_i/u_i$. Entrywise,
\[
 B(u,v)=U\diag(u_i^{-1})B_1(z)\diag(u_i^{-1}),
 \qquad \Delta(u,v)=-4U^2q(z).
\]
Thus $\Delta$ is irreducible after inverting the $u_i$, apart from a unit. No $u_i$ divides $\Delta$, since
\begin{equation}\label{eq:boundarydelta}
 \left.\Delta\right|_{u_i=0}
 =-4v_i^2\prod_{j\ne i}u_j
   \left(\sum_{j\ne i}u_j\prod_{k\ne i,j}v_k\right)
\end{equation}
is a nonzero polynomial. Therefore there is no factor supported on the omitted coordinate boundary, and $\Delta$ is irreducible in $\C[u,v]$. Applying the same change of variables over $\C(u)$ proves the last assertion.
\end{proof}

\begin{lemma}\label{lem:tenranktwo}
The nonzero vectors $a$ for which $\rk B_1(a)\leq2$ are exactly the ten lines spanned by the four permutations of $(-1,1,1,1)$ and the six vectors $e_i-e_j$ with $i<j$. Every matrix on these lines has rank exactly two.
\end{lemma}
\begin{proof}
A nonzero symmetric matrix of rank one is a nonzero scalar multiple of $vv^T$. It cannot have zero diagonal, since that would force every $v_i^2$ to vanish. A symmetric matrix of rank two and zero diagonal can be written $rs^T+sr^T$: diagonalize its quadratic form of rank two and replace the two diagonal basis vectors by suitable sum and difference vectors. The zero diagonal says $r_i s_i=0$ for every $i$. The nonzero pattern is therefore complete bipartite between two disjoint supports.

No row of a nonzero matrix $B_1(a)$ of rank at most two is zero. If row $i$ were zero, the three pair sums among the other coordinates of $a$ would vanish. Those three coordinates would be zero. The determinant of the complementary $3\times3$ block would then be $2a_i^3$, forcing $a_i=0$ too.

The bipartition thus uses all four indices. For a $1+3$ partition with center $1$, the three zero edges among the other indices give $a_1+a_2=a_1+a_3=a_1+a_4=0$. This is the first listed star line. For a $2+2$ partition $\{1,2\}\mid\{3,4\}$, the two internal zero edges give $a=(p,-p,q,-q)$. The determinant of the cross $2\times2$ block is $-4pq$. Rank two of the full matrix requires that cross block to have rank one, so $p=0$ or $q=0$. These are the pair difference lines. Permuting the indices gives all ten lines. Direct substitution gives rank two on each, proving both inclusion and completeness.
\end{proof}

\subsection{The singular locus}\label{sec:singular}
The partial derivative $\partial f/\partial x_{ij}$ is the permanent of the $3\times3$ submatrix that omits row $i$ and column $j$. Expanding $f$ along a row writes it as a combination of these subpermanents. Therefore the singular locus $\Sing(f)$ of the hypersurface $f=0$ is exactly the set where all sixteen $3\times3$ subpermanents vanish. Its dimension is eight. This is also recorded in~\cite{ABV}, and the first version of this paper obtained it from a Gr\"obner basis computed with Singular~\cite{Singular}. The proof given here needs no computer. It starts with a lemma on $2\times2$ permanents.

\begin{lemma}\label{lem:zero2}
A matrix all of whose $2\times2$ permanents vanish is supported in one row, in one column, or in one $2\times2$ rectangle on which $ad+bc=0$.
\end{lemma}
\begin{proof}
If one row has at least three nonzero entries, divide another row's entries by those entries on three occupied columns. Vanishing of the mixed $2\times2$ permanents gives $t_i+t_j=0$ for every pair. Since $2\ne0$, all three ratios are zero. Comparing any remaining column with an occupied column then shows that the other row is entirely zero. Thus only the original row can be nonzero.

If one row has exactly two nonzero entries $a,b$, every other row vanishes outside those two columns, by comparison with either occupied entry. Its entries in those columns must be $(\lambda a,-\lambda b)$. Two nonzero additional rows would have permanent $-2\lambda\mu ab\ne0$ on that rectangle. At most one additional row is therefore nonzero.

Finally, if every row has at most one nonzero entry, two entries in different rows and different columns would have a nonzero $2\times2$ permanent. All entries must lie in one column, unless they already lie in one row. This covers every possibility.
\end{proof}
For a $3\times3$ matrix the set in Lemma~\ref{lem:zero2} has dimension three: its row and column spaces have dimension three, and each $2\times2$ quadric cone has dimension three. Since these are precisely the gradient equations of $\per_3$, we obtain $\dim\Sing(\per_3)=3$.

\begin{lemma}\label{lem:singF}
The singular locus of the cubic $F(a,b,c)=\per_4(\mathbf1,a,b,c)$ has dimension four.
\end{lemma}
\begin{proof}
Its equations are
\[
 B_1(a)b=0,\qquad B_1(a)c=0,\qquad B_1(b)c=0.
\]
We stratify by $a$. If $B_1(a)$ has rank four, then $b=c=0$, and this stratum has dimension four.

If it has rank three, write its kernel as $\C k$. Then $b=sk$ and $c=tk$, and the remaining equations become $stB_1(k)k=0$. The exceptional condition $B_1(k)k=0$ holds only when $k$ is a coordinate vector. Indeed its $i$th entry is twice the second elementary symmetric function of the three coordinates other than $i$. A vector of support two or three makes one of these expressions a nonzero product. If all four coordinates are nonzero, division by the product of the three coordinates gives
\[
 \sum_{j\ne i}\frac1{k_j}=0\quad\text{for every }i.
\]
Writing $T=\sum_j1/k_j$, this says $1/k_i=T$ for all $i$, and hence $T=4T$, an impossibility. Only support one remains.

If $k=e_i$, the equation $B_1(a)e_i=0$ forces all $a_j$ with $j\ne i$ to vanish. These exceptional $a$ form four lines, so even their two dimensional $(s,t)$ fibers give dimension at most three. For every other $a$ of rank three, the condition is $st=0$, a fiber of dimension one. The base of rank three has dimension at most three because $\det B_1$ is a nonzero polynomial. Its contribution is at most four.

If $B_1(a)$ has rank two, Lemma~\ref{lem:tenranktwo} gives a base of dimension one. Both $b$ and $c$ belong to its two dimensional kernel $L$. The equation $B_1(b)c=0$ is not identically zero on $L\times L$: otherwise $B_1(w)w=0$ for every $w\in L$, whereas the preceding paragraph showed that this zero set is a finite union of coordinate lines and cannot contain a plane. A nonzero polynomial on the four dimensional affine space $L\times L$ cuts its dimension to at most three. This stratum has dimension at most four.

There is no nonzero case of rank one. In rank zero, injectivity of $a\mapsto B_1(a)$ gives $a=0$. We are left with $B_1(b)c=0$. For ranks four, three, two, and zero of $B_1(b)$ the dimensions of the corresponding incidence strata are respectively bounded by
\[
 4+0,
 \quad 3+1,
 \quad 1+2,
 \quad 0+4.
\]
They are all at most four. The original stratum $b=c=0$ already has dimension four, proving equality.
\end{proof}

\begin{proposition}\label{prop:singdim}
The singular locus of $f=\per_4$ has dimension eight in $\C^{16}$.
\end{proposition}
\begin{proof}
Every partial derivative of $f$ is a $3\times3$ subpermanent. Euler's identity shows that their simultaneous vanishing also implies $f=0$. Matrices with two zero rows form an eight dimensional subspace of the singular locus, so the dimension is at least eight.

For the reverse inequality, cover the nonzero matrices by the finitely many choices of a nonzero row and its support. Permute that row to the first position and call it $z$. The equations obtained by differentiating in the other three rows say that those rows belong to $\Sing(F_z)$. Additional equations from differentiation in the first row can only reduce the fiber dimension.

If $z$ has support four, column scaling reduces $F_z$ to a nonzero scalar multiple of $F_{\mathbf1}$, with an invertible change in each lower row. Lemma~\ref{lem:singF} gives fiber dimension four. The first row varies in dimension four, giving a total of at most eight.

If $z$ has support one, $F_z$ is a nonzero scalar multiple of a $3\times3$ permanent in nine essential variables, independent of one coordinate in each lower row. Thus $\dim\Sing(F_z)=3+3=6$, and the total dimension is at most $1+6=7$.

If $z$ has support two, suppose it is $(z_1,z_2,0,0)$. In each lower row replace the first two coordinates by
\[
 t=z_2y_1+z_1y_2
\]
and an independent complementary coordinate. Then $F_z$ is the $3\times3$ permanent of the three lower rows with coordinates $(t,y_3,y_4)$. It is independent of the three complementary coordinates. Again the singular fiber has dimension six, and the first row contributes two, giving at most eight.

It remains to consider support three. By column scaling, take $z=(1,1,1,0)$. Write each lower row as $(v_i,t_i)$ with $v_i\in\C^3$ and $t_i\in\C$, and set $C=C(1,1,1)$. The contracted cubic is
\[
 t_1v_2^TCv_3+t_2v_1^TCv_3+t_3v_1^TCv_2.
\]
The matrix $C$ is nonsingular. Its gradient equations in the $v_i$ are therefore equivalent to
\[
 t_2v_3+t_3v_2=0,\quad
 t_1v_3+t_3v_1=0,\quad
 t_1v_2+t_2v_1=0,
\]
and the derivatives in the $t_i$ require the three pairwise products $v_i^TCv_j$ to vanish.

If all three $t_i$ are nonzero, the displayed linear equations force all $v_i=0$, giving dimension three. If exactly two $t_i$ are nonzero, one $v_i$ is zero and the other two are proportional. Their pairwise orthogonality requires a nondegenerate quadratic equation on a vector in $\C^3$, so the dimension is $2+2=4$. If exactly one $t_i$ is nonzero, the other two vectors vanish and the remaining vector is arbitrary, again giving dimension $1+3=4$.

Suppose all $t_i=0$. The three vectors must be pairwise $C$-orthogonal. Differentiation of the original permanent in $x_{14}$ supplies the further equation
\[
 \per_3(v_1,v_2,v_3)=0.
\]
The locus of linearly dependent pairwise orthogonal triples has dimension at most five. If one vector is zero, the other two satisfy one nonzero bilinear equation in six variables. If no vector is zero and the span has dimension two, a dependence and pairwise orthogonality force at least two of the vectors to be isotropic and proportional: a nondegenerate three dimensional quadratic space has maximal isotropic dimension one. Choose that isotropic line (one parameter), the two vectors on it (two parameters), and the third vector in its orthogonal plane (two parameters). The bound is five. Triples of rank at most one give no larger dimension.

The pairwise orthogonal triples of full rank form an irreducible variety of dimension six. After a constant change making $C$ the identity, they are the image of
\[
 \mathrm{SO}_3(\C)\times(\C^*)^3,
 \qquad (Q,s_1,s_2,s_3)\longmapsto Q\diag(s_1,s_2,s_3).
\]
Surjectivity follows by taking square roots of the three nonzero norms and choosing their signs so the normalized orthogonal matrix has determinant one. The group $\mathrm{SO}_3(\C)$ is irreducible of dimension three; this also follows from its Cayley parametrization by skew $3\times3$ matrices on a dense open subset. Hence the displayed image is irreducible of dimension six.

The permanent is not identically zero on that image. A witness, with the vectors as rows, is
\[
 O=\begin{pmatrix}1&1&2\\7&-3&-6\\1&-3&3\end{pmatrix},
 \qquad OCO^T=\diag(10,-90,-18),\quad
 \det O=-90,\quad \per_3(O)=-24.
\]
The extra permanent equation therefore reduces the dimension of this irreducible six dimensional locus to at most five. Together with the triples of lower rank, the fiber over $t=0$ has dimension at most five. Adding the three nonzero coordinates of the first row gives at most $3+5=8$.

Every nonzero first row has now been covered, and after a row permutation so has every nonzero matrix. Thus the singular locus has dimension at most eight.
\end{proof}

\subsection{Regularity and the normal form}\label{sec:normalform}
\begin{lemma}\label{lem:regularity}
If $A$ is a representation and $f(p)=0$, then $\rk A(p)=m-1$.
\end{lemma}
\begin{proof}
Let $\mathcal D$ be the set of $m\times m$ matrices of rank at most $m-2$, and let $Z=A^{-1}(\mathcal D)$ be its preimage. The matrices of rank exactly $m-2$ form an irreducible set of dimension $(m-2)(m+2)$, which is dense in $\mathcal D$, so $\mathcal D$ is irreducible of codimension $4$. Hence the codimension of every component of $Z$ in $\C^{16}$ is at most $4$ for nonempty $Z$. The gradient of $\det$ at a matrix of rank at most $m-2$ is zero, since its entries are $(m-1)$-minors. By the chain rule the gradient of $f$ vanishes on $Z$, so $Z\subseteq\Sing(f)$. Proposition~\ref{prop:singdim} gives codimension $8$ for the singular locus, so $Z$ is empty. Compare~\cite{vzG}.
\end{proof}
This is also the regularity argument of~\cite[Proposition~2.1]{ABV}.

Note that since $f(0)=0$, the matrix $A(0)$ has rank $m-1$ by Lemma~\ref{lem:regularity}. Multiply $A$ on both sides by constant invertible matrices, and then rescale one row. This enables us to assume that $A(0)=\diag(0,I_h)$ with $h=m-1$. This gives
\[
 A(X)=\begin{pmatrix}\ell(X)&-b(X)\\c(X)&I_h-D(X)\end{pmatrix},
\]
where $\ell$ is a linear form, $b$ is a row of $h$ linear forms, $c$ is a column of $h$ linear forms and $D$ is an $h\times h$ matrix of linear forms, all of them homogeneous.

\begin{lemma}\label{lem:normalform}
$\ell=0$, and
\begin{equation}\label{eq:lowmoments}
 bc=0,\qquad bDc=0,\qquad bD^2c=f.
\end{equation}
\end{lemma}
\begin{proof}
The Schur complement gives $\det A=\ell\det(I-D)+b\adj(I-D)c$. As formal power series,
\begin{equation}\label{eq:schur}
 \adj(I-D)=\det(I-D)(I-D)^{-1}
 =\left(\sum_{i=0}^h(-1)^i e_i(D)\right)\left(\sum_{j\geq0}D^j\right),
\end{equation}
where $e_i(D)$ is the sum of the principal $i\times i$ minors of $D$ and is homogeneous of degree $i$. Now compare homogeneous parts with $f$, which is homogeneous of degree $4$. Degree $1$ gives $\ell=0$. Degrees $2$, $3$ and $4$ then give $bc=0$, $bDc-e_1(D)bc=0$ and $bD^2c-e_1(D)bDc+e_2(D)bc=f$.
\end{proof}
In general, with $\mu_j=bD^jc$, the comparison in degree $d$ reads
\[
 \sum_{i+j=d-2}(-1)^i e_i(D)\mu_j=[f]_d.
\]

\subsection{The endpoint inequality}\label{sec:endpoint}
Put $H=\C^h$. The endpoint map is
\[
 \Phi:\C^{16}\longrightarrow W=H^*\oplus H,\qquad \Phi(y)=(b(y)^T,c(y)).
\]
Its kernel $K$ is the endpoint kernel of the representation, its image is $U$, and $a=\dim U=16-\dim K$. We give $W$ the nondegenerate symmetric form
\[
 \langle(\beta,\gamma),(\beta',\gamma')\rangle
 =\beta\gamma'+\beta'\gamma,
\]
and for $z\in\C^{16}$ we put $M_z=\diag(D(z)^T,D(z))$. Each $M_z$ is self-adjoint for this form. We write $\operatorname{Hess}_z(f)$ for the Hessian of $f$ at $z$, viewed as a symmetric bilinear form on $\C^{16}$, $\rho(z)$ for its rank, and $\rho_K$ for the largest value of $\rho$ on $K$, which is its value at a generic point of $K$. Finally put $E=U^\perp/U$. The proof below shows that $U$ is isotropic, so the form on $W$ induces a nondegenerate form on $E$, and $\dim E=2h-2a$.

\begin{proposition}\label{prop:endpoint}
$K\subseteq\Sing(f)$, and $\rho(z)\leq2(h-a)$ for every $z\in K$. Therefore
\begin{equation}\label{eq:endpointbound}
 \rho_K\leq2(h-a),\qquad
 m\geq17-\dim K+\left\lceil\frac{\rho_K}{2}\right\rceil.
\end{equation}
\end{proposition}
\begin{proof}
Let $T(s_1,s_2,s_3,s_4)=b(s_1)D(s_2)D(s_3)c(s_4)$, so that $f(x)=T(x,x,x,x)$. Define the trilinear form $R(s_1,s_2,s_3)=b(s_1)D(s_2)c(s_3)$ in the same way. For $z\in K$, every value of $T$ or $R$ with $z$ in the first or last slot is zero. In particular $f(z)=0$. Now take the derivative of $f$ at $z$ in a direction $y$. It is a sum of four values of $T$, each with $z$ in three of the four slots. In each of them $z$ occupies the first or the last slot, so each value is zero. Hence $K$ lies in the singular locus.

Polarizing $bc=0$ gives $\langle\Phi(y),\Phi(y')\rangle=b(y)c(y')+b(y')c(y)=0$, so $U$ is isotropic and $U\subseteq U^\perp$. Polarizing $bDc=0$ at $(y,z,y')$ with $z\in K$ leaves only the two orderings with $z$ in the middle. Hence $b(y)D(z)c(y')+b(y')D(z)c(y)=0$. The left side is $\langle M_z\Phi(y),\Phi(y')\rangle$, so $M_zU\subseteq U^\perp$.

Note that $f$ is homogeneous of degree $4$, so $\operatorname{Hess}_z(f)(y,y')=\frac12\partial_z\partial_z\partial_y\partial_{y'}f$. Also, the fourth derivative is the sum of $T$ over the $24$ orderings of $(y,y',z,z)$. Only the four orderings with $y$ and $y'$ at the ends survive. Therefore, writing $R_z(y)=M_z\Phi(y)+U\in E$,
\begin{equation}\label{eq:gramhess}
 \operatorname{Hess}_z(f)(y,y')
 =b(y)D(z)^2c(y')+b(y')D(z)^2c(y)
 =\langle R_z(y),R_z(y')\rangle_E.
\end{equation}
The map $R_z$ is linear into $E$, and $\operatorname{Hess}_z(f)$ is the pullback of the form on $E$ along it. So $\rho(z)\leq2h-2a$, which is $h\geq16-\dim K+\rho(z)/2$.
\end{proof}
When equality holds in the rank bound, $R_z$ is surjective for generic $z$ and
\begin{equation}\label{eq:tightkernel}
 \ker R_z=\operatorname{rad}\operatorname{Hess}_z(f),
\end{equation}
because a surjective map to a nondegenerate space has pullback radical equal to its kernel.

The inequality uses the Hessian in the spirit of~\cite{MR}, but at a point of the endpoint kernel rather than at a generic zero. It is weakest when $\rho_K=0$, which happens when $K$ is spanned by a row.

\subsection{Moments on the endpoint kernel}\label{sec:moments}
At a point $z\in K$ we have $A(z)=\diag(0,I-D(z))$. Lemma~\ref{lem:regularity} says that $\det(I-D(z))$ never vanishes for $z\in K$. A nonconstant complex polynomial has a zero, by restriction to an affine line on which it is nonconstant. Since the value at $z=0$ is one, we obtain
\begin{equation}\label{eq:nilpotence}
 \det(I-D(z))=1\quad(z\in K),\qquad D(z)\text{ is nilpotent}.
\end{equation}
For the second assertion replace $z$ by $tz$: the characteristic polynomial has no nonzero coefficient below its leading term.

Choose a complement to $K$ and write its variables as $y$; we call them the endpoint variables. The part of the determinant of degree two in $y$ is
\[
 b(y)(I-D(z))^{-1}c(y).
\]
The part of $f$ of degree two in $y$ has degree two in $z$. Comparing powers of $z$, and polarizing in $y$, therefore yields
\begin{equation}\label{eq:allmoments}
 \langle U,M_z^jU\rangle=0\quad(j\ne2),
 \qquad
 \langle\Phi(y),M_z^2\Phi(y')\rangle
 =\operatorname{Hess}_z(f)(y,y').
\end{equation}

\begin{lemma}\label{lem:cubicresolvent}
Suppose $z\in K$, $M_zU\subseteq U$, and the Hessian at $z$ is zero. Define $S$ on the endpoint variables by $M_z\Phi(y)=\Phi(Sy)$; this uses that $\Phi$ is injective on the chosen complement. If the part of $f(tz+y)$ of degree three in $y$ is $tP(y)$, then
\[
 S\text{ is nilpotent},\qquad
 P(y)=-b(y)D(Sy)c(y),\qquad
 P((I-tS)^{-1}y)=P(y),\qquad \delta_SP=0.
\]
Here $\delta_SP$ denotes the derivative of $P$ in the linear vector field $Sy$.
\end{lemma}
\begin{proof}
Nilpotence follows by restricting the nilpotent operator $M_z$ to its invariant subspace. Write $R=(I-tD(z))^{-1}$. In the part of the Schur complement of degree three in $y$, the term coming from the factor $\det(I-D)$ is a multiple of the part $bRc$ of degree two, which is zero because the Hessian at $z$ is zero and \eqref{eq:allmoments} holds. The cubic part is therefore
\[
 b(y)R D(y)R c(y)=tP(y).
\]
Intertwining gives $b(y)R=b((I-tS)^{-1}y)$ and $Rc(y)=c((I-tS)^{-1}y)$. Put $y'=(I-tS)^{-1}y$. Since $y=y'-tSy'$ and $b(y')D(y')c(y')=0$, the left side equals
\[
 -t\,b(y')D(Sy')c(y').
\]
Its coefficient of $t$ at zero identifies $P(y)=-b(y)D(Sy)c(y)$. Substituting this in the whole identity gives $P((I-tS)^{-1}y)=P(y)$. Differentiating at $t=0$ proves $\delta_SP=0$.
\end{proof}

A representation of size $m<14$ gives one of size $14$ by adjoining an identity block, with the same endpoint kernel. So Theorem~\ref{thm:main} follows from Grenet's construction and one remaining claim. Hence, the claim that no representation of size $14$ exists is all that is left to prove. From now on we set
\begin{equation}\label{eq:size14}
 m=14,\quad h=13,\quad k=\dim K,\quad a=16-k,
 \quad \dim E=2(k-3),\quad \rho_K\leq2(k-3).
\end{equation}

\section{Admissible kernels}
\label{sec:classification}
The endpoint inequality turns the question into one about linear spaces inside the singular locus. This section lists the admissible ones. With $m=14$, Section~\ref{sec:endpoint} gives
\begin{equation}\label{eq:admissible}
 K\subseteq\Sing(f),\qquad k\geq3,\qquad \rho_K\leq2(k-3),
 \qquad \dim E=2(k-3).
\end{equation}
We call a linear subspace with the first three properties admissible. Limits of subspaces are used below only to solve this geometric problem. A limit of endpoint kernels is never claimed to be the endpoint kernel of a representation.

Two groups act. The group $G$ of order $1152$ generated by permutations of rows, permutations of columns and transposition preserves $f$. The torus $(\C^*)^8$ acts by $x_{ij}\mapsto s_it_jx_{ij}$ and multiplies $f$ by $\prod_is_i\prod_jt_j$. Let $g$ be an element of either group. Substitute $g$ into a representation, and rescale one row. The result is another representation of the same size. Its endpoint kernel is the original kernel, transformed by $g^{-1}$. Both groups preserve $\Sing(f)$ and Hessian ranks.

\subsection{Coordinate kernels}
Let $K_S$ be the span of the variables $x_{ij}$ with $(i,j)\in S$, for a set $S$ of cells. The restriction of $\partial f/\partial x_{ij}$ to $K_S$ is a sum of distinct monomials. Each monomial corresponds to one permutation through $(i,j)$ whose other three cells lie in $S$. Hence $K_S\subseteq\Sing(f)$ exactly when every permutation has at most two of its cells in $S$.

\begin{proposition}\label{prop:coordinate}
Up to $G$ there are $27$ coordinate subspaces of $\Sing(f)$, counting the zero space. Exactly nine are admissible. They are listed in Table~\ref{tab:coordinate}.
\end{proposition}
\begin{proof}
Enumerate all $2^{16}=65{,}536$ sets $S$, keep those that meet every permutation in at most two cells, and compute $\rho$ as the rank of the Hessian with the variables outside $S$ set to zero, over the field of rational functions. Exactly $3{,}017$ sets pass the first test, and they form $27$ orbits under $G$. In practice $S$ is a sixteen bit integer and each permutation is the bit mask of its four cells; $S$ is rejected when it shares more than two bits with one of the twenty-four masks, and the smallest mask in each orbit represents the orbit. The Hessian entries come from differentiating the sum of the twenty-four monomials of $f$. After the variables outside $S$ are set to zero, the rank is found by exact Gaussian elimination over $\Q(x_s:s\in S)$, so no probabilistic rank estimate is involved. The program \file{k6_attempt/coordinate_kernels_checks.py} in archive~A carries out this computation.
\end{proof}

\begin{table}[ht]
\centering\small
\renewcommand{\arraystretch}{1.3}
\begin{tabular}{lcrrr}\toprule
Kernel & Cells & $\dim K$ & $\rho_K$ & bound\\\midrule
two rows &
$\left[\begin{smallmatrix}\bullet&\bullet&\bullet&\bullet\\\bullet&\bullet&\bullet&\bullet\\\cdot&\cdot&\cdot&\cdot\\\cdot&\cdot&\cdot&\cdot\end{smallmatrix}\right]$
&8&8&13\\
full $2\times3$ block &
$\left[\begin{smallmatrix}\bullet&\bullet&\bullet&\cdot\\\bullet&\bullet&\bullet&\cdot\\\cdot&\cdot&\cdot&\cdot\\\cdot&\cdot&\cdot&\cdot\end{smallmatrix}\right]$
&6&4&13\\
one row &
$\left[\begin{smallmatrix}\bullet&\bullet&\bullet&\bullet\\\cdot&\cdot&\cdot&\cdot\\\cdot&\cdot&\cdot&\cdot\\\cdot&\cdot&\cdot&\cdot\end{smallmatrix}\right]$
&4&0&13\\
two rows less a cell &
$\left[\begin{smallmatrix}\bullet&\bullet&\bullet&\bullet\\\bullet&\bullet&\bullet&\cdot\\\cdot&\cdot&\cdot&\cdot\\\cdot&\cdot&\cdot&\cdot\end{smallmatrix}\right]$
&7&8&14\\
row and two cells &
$\left[\begin{smallmatrix}\bullet&\bullet&\bullet&\bullet\\\bullet&\bullet&\cdot&\cdot\\\cdot&\cdot&\cdot&\cdot\\\cdot&\cdot&\cdot&\cdot\end{smallmatrix}\right]$
&6&6&14\\
$K_6$ &
$\left[\begin{smallmatrix}\bullet&\bullet&\bullet&\bullet\\\bullet&\cdot&\cdot&\cdot\\\bullet&\cdot&\cdot&\cdot\\\cdot&\cdot&\cdot&\cdot\end{smallmatrix}\right]$
&6&6&14\\
$2\times3$ block less a cell &
$\left[\begin{smallmatrix}\bullet&\bullet&\bullet&\cdot\\\bullet&\bullet&\cdot&\cdot\\\cdot&\cdot&\cdot&\cdot\\\cdot&\cdot&\cdot&\cdot\end{smallmatrix}\right]$
&5&4&14\\
cross with arms of two &
$\left[\begin{smallmatrix}\bullet&\bullet&\bullet&\cdot\\\bullet&\cdot&\cdot&\cdot\\\bullet&\cdot&\cdot&\cdot\\\cdot&\cdot&\cdot&\cdot\end{smallmatrix}\right]$
&5&4&14\\
row less a cell &
$\left[\begin{smallmatrix}\bullet&\bullet&\bullet&\cdot\\\cdot&\cdot&\cdot&\cdot\\\cdot&\cdot&\cdot&\cdot\\\cdot&\cdot&\cdot&\cdot\end{smallmatrix}\right]$
&3&0&14\\\bottomrule
\end{tabular}
\caption{The admissible coordinate kernels up to symmetry. The last column is the lower bound $17-\dim K+\lceil\rho_K/2\rceil$ of Proposition~\ref{prop:endpoint}. The complete cross, a row together with a column, lies in the singular locus but forces size $15$.}
\label{tab:coordinate}
\end{table}

The table shows that each admissible coordinate kernel lies in one of three containers: a single row, two rows, or the complete cross spanned by row $1$ and column $1$. A cross with arms of lengths $p$ and $q$ consists of one cell, its center, together with $p$ further cells in its row and $q$ further cells in its column, so it has dimension $1+p+q$. Thus $K_6$ is the cross with arms of lengths three and two, and the cross with arms of two has dimension five.

\subsection{Kernels with zero Hessian}
If a $2\times2$ permanent $p$ in rows $i,j$ and columns $r,s$ is nonzero at a point, consider the four Hessian indices in the complementary $2\times2$ rectangle. The corresponding principal submatrix is
\[
 \begin{pmatrix}0&0&0&p\\0&0&p&0\\0&p&0&0\\p&0&0&0\end{pmatrix},
 \qquad \det=p^4.
\]
Thus a nonzero Hessian has rank at least four. For $k=3$ or $4$, condition \eqref{eq:admissible} permits rank at most zero or two, respectively. All $2\times2$ permanents must therefore vanish identically on $K$.

By Lemma~\ref{lem:zero2}, the locus defined by the $2\times2$ permanents is the finite union of the coordinate row spaces, coordinate column spaces, and the quadrics $ad+bc=0$ on coordinate $2\times2$ rectangles. A linear space is irreducible, so if it lies in this finite union, it lies in one member. A linear subspace of the nondegenerate quadric $ad+bc=0$ has dimension at most two: its polarization is a nondegenerate symmetric form on a four dimensional space, and a totally isotropic subspace has dimension at most half that dimension. Consequently an admissible $K$ of dimension three or four lies in a single row or column.

A three dimensional subspace of a row is a hyperplane. Its defining covector can, by independent column scalings, be normalized to have all its nonzero entries equal to one. Up to column permutations there are four types, according to the number of nonzero entries of the covector. Dimension four gives the whole row. This argument covers noncoordinate row hyperplanes as well as coordinate ones.

\subsection{Torus limits}
Let $X_k\subseteq\Gr(k,\C^{16})$ be the set of admissible $k$-planes. It is closed. In a Grassmannian chart, write a general point of a plane using $k$ independent parameters. The coefficients of the sixteen $3\times3$ permanents must all vanish; the coefficients of every Hessian minor of size $2(k-3)+1$ must also vanish. These are polynomial equations in the chart coordinates. They define a closed condition independently of the chart.

The torus $(\C^*)^8$ preserves $X_k$. An explicit one-parameter subgroup has row weights $0,10,20,30$ and column weights $0,1,2,3$. Its sixteen cell weights are all distinct. Order the coordinates by increasing weight and put a basis of $K$ into row echelon form. For each basis vector, its first nonzero coordinate is a distinct pivot and every remaining nonzero coordinate has greater weight. Apply the one-parameter subgroup and divide each basis vector by the power of the parameter belonging to its pivot. As the parameter tends to zero, that vector tends to the corresponding coordinate vector. Thus the Grassmannian limit $K_0$ is a coordinate subspace of the same dimension. Closedness gives $K_0\in X_k$.

So $3\leq k\leq8$, and $K_0$ is one of the kernels in Table~\ref{tab:coordinate}. To recover information about $K$ itself we need more than the existence of this limit. Suppose a neighborhood of $K_0$ inside $X_k$ consists of planes contained in one fixed space of two rows or one fixed cross. The orbit map, extended to the parameter value zero, pulls this neighborhood back to an open set containing zero. It therefore contains a nonzero parameter. The corresponding torus translate of $K$ lies in the container. The inverse torus transformation preserves that coordinate container, so $K$ itself lies in it.

We now prove this local statement. A first order calculation alone would not be enough, since a curved branch could have all its tangent directions inside a container; the argument after Table~\ref{tab:charts} deals with this.

\subsection{Local containment}\label{sec:charts}
Fix one of the nine coordinate limits. Write a nearby plane as the graph of a linear map from that coordinate plane to its coordinate complement. Separate its graph coefficients into $p$, the coefficients inside the proposed container, and $q$, the coefficients outside it. The coefficients of every $3\times3$ permanent vanish identically at $q=0$, for every value of $p$, because the whole container lies in $\Sing(f)$.

We also use Hessian minors, chosen so that they vanish identically when $q=0$, for every $p$. For a container of two rows, every row of the Hessian indexed by a variable of an occupied row is zero. After differentiating in that variable, a surviving term of degree two would still require a variable from an empty row. For a cross, every row of the Hessian indexed by a variable of the cross is likewise zero: differentiating in the row arm leaves too many distinct columns to be supplied from the single column arm, and the statement for the column arm is its transpose.

At a point $z$ of the coordinate plane, select an invertible principal block $P=H_{I,I}$ of the Hessian of size $r=2(k-3)$, in the cases where the generic Hessian has that rank. Append one of the zero rows $i$ just described and any column $j$ to obtain an $(r+1)$-minor. Its derivative in a graph direction is
\begin{equation}\label{eq:chartminor}
 \det(P)\bigl(dH_{ij}-dH_{i,I}P^{-1}H_{I,j}\bigr).
\end{equation}
This follows by differentiating the Schur complement formula for the enlarged determinant at a point where its last row is zero. These minors are polynomial equations that vanish on $X_k$, and they vanish identically when $q=0$.

Take the coefficients of the singularity equations together with the derivatives~\eqref{eq:chartminor}. Table~\ref{tab:charts} gives the ranks of their Jacobian with respect to $q$; the last column is a full size minor, taken after constant row normalizations.
\begin{table}[ht]
\centering\small
\begin{tabular}{lrrrr}\toprule
Coordinate limit & $\#q$ & rank, singularity equations & rank, with minors & minor\\\midrule
$2\times3$ block less a cell &40&36&40&$-64$\\
row and two cells &48&46&48&$16$\\
cross with arms of two &45&39&45&$-26880$\\
$K_6$ &54&51&54&$-128$\\\bottomrule
\end{tabular}
\caption{Ranks of the Jacobians in the local containment argument.}
\label{tab:charts}
\end{table}
In the first, second and fourth cases all free variables of the coordinate plane are set equal to one. For the cross with arms of two that point gives rank $43$, and the point $(s,a,b,d,e)=(1,2,3,5,7)$ raises it to $45$. The indices of the selected minors, the evaluation points and the row normalizations are stored in \file{k6_attempt/chart_jacobian_certificate.json} in archive~A, and its checker recomputes the derivatives.

To turn this into a local statement, select $\#q$ equations $F(p,q)$ whose Jacobian minor is the nonzero minor in the table. Since $F(p,0)=0$ identically, each monomial of each component contains a $q$ variable. Factor one such variable from each monomial, obtaining a polynomial matrix $A(p,q)$ with
\[
 F(p,q)=A(p,q)q,\qquad A(0,0)=\frac{\partial F}{\partial q}(0,0).
\]
Its determinant is nonzero at the origin. On the open set where it remains nonzero, $F=0$ implies $q=0$, so a whole neighborhood of $K_0$ in $X_k$ lies in the container.

The three remaining coordinate limits with nonzero Hessian contain a full $2\times3$ block: the full $2\times3$ block itself, two rows less a cell, and two rows. Denote the entries of their first three columns by $u_1,u_2,u_3$ and $v_1,v_2,v_3$. If $w$ is a row outside the container, in a graph plane, then singularity on those columns gives
\[
 w_1(u_2v_3+u_3v_2)+w_2(u_1v_3+u_3v_1)
 +w_3(u_1v_2+u_2v_1)=0.
\]
The three quadrics have no syzygy with linear coefficients: the coefficient map has rank $18$ on its $18$ unknowns (Section~\ref{sec:syzygies}). The same conclusion holds with additional independent variables: their coefficients would supply constant syzygies, and the three quadrics are linearly independent. Thus $w_1=w_2=w_3=0$. Singularity on columns $1,2,4$ then gives $w_4(u_1v_2+u_2v_1)=0$, hence $w_4=0$ in the polynomial ring. This holds on the whole chart. The coefficient maps for the outside rows have ranks $24$, $28$ and $32$ in the three cases, and \file{k6_attempt/grassmann_chart_checks.py} in archive~A recomputes them.

The cases with zero Hessian were handled without charts. Together with the torus argument, this proves the following.
\begin{proposition}\label{prop:containers}
Every admissible endpoint kernel lies, up to row and column permutations and transposition, in two coordinate rows or in a complete coordinate row--column cross.
\end{proposition}

\subsection{Kernels inside two rows}\label{sec:insidetworows}
Write the two rows as $u,v\in\C^4$. The Hessian on the other two rows is
\[
 \begin{pmatrix}0&B(u,v)\\B(u,v)^T&0\end{pmatrix}.
\]
All other Hessian rows vanish. Thus if $r$ is the generic rank of $B$ on $K$, admissibility is exactly $r\leq k-3$.

We will also use two consequences of the explicit matrix $B$. For every fixed nonzero $v$, the generic rank as a function of $u$ is at least three. Indeed, choose $i$ with $v_i\ne0$, set $u_i=0$, and take the complementary principal $3\times3$ minor. It is
\[
 2v_i^3\prod_{j\ne i}u_j\ne0.
\]
Furthermore, $\Delta(u,v)$ vanishes identically in $u$ precisely when $v$ has support at most two. For support at most two one checks a fixed kernel directly. For support three or four, normalize its nonzero entries by column scaling. With $u=\mathbf1$, the two representatives $v=(1,1,1,0)$ and $v=\mathbf1$ give determinants $-12$ and $-48$, respectively.

\paragraph{Dimension five.}
Suppose first that the projection to the $u$ row contains a vector of full support. Let that projection have dimension $s\leq4$, and let $V_0$ be its kernel, of dimension $5-s$. For generic $u$, the fiber of $K$ is an affine space $v_0+V_0$. After scaling $u$ to $\mathbf1$, the condition $\rk B\leq2$ restricts $v$ to the ten lines of Lemma~\ref{lem:tenranktwo}. An affine space is irreducible and hence lies in one of those lines. Therefore $5-s\leq1$, forcing $s=4$ and $\dim V_0=1$. An affine line contained in a line through zero is that entire line, so every generic fiber is $V_0$ itself. It follows that $(u,0)\in K$ generically and hence identically, and $K=\C^4\oplus V_0$. This contradicts the rank bound above for any nonzero $v\in V_0$.

So neither row projection contains a vector of full support. Each is contained in one coordinate hyperplane, because a linear space cannot be covered by finitely many proper linear subspaces. Their dimensions sum to at least five, so one projection is a three dimensional coordinate hyperplane, say $u_4=0$. If $v_4$ were not identically zero on $K$, the polynomial $2v_4^3u_1u_2u_3$ would be nonzero on the integral coordinate ring of $K$, contradicting rank at most two. Hence $v_4=0$ also. Thus $K$ is a hyperplane in a full $2\times3$ block. Conversely every such hyperplane has $r\leq2$ and is admissible.

\paragraph{Dimension six.}
Suppose the $u$ projection contains a vector of full support, and again denote its dimension by $s$. Its generic fibers have dimension $6-s\geq2$ and lie in $\Delta(u,v)=0$. For fixed $u$ of full support, this is an irreducible homogeneous quartic hypersurface in four $v$ variables by Lemma~\ref{lem:crossdet}. It contains no affine three-plane: a three-plane is an affine hyperplane, whose linear equation would divide the quartic. If an affine two-plane $v_0+V_0$ did not contain zero, homogeneity would make its three dimensional linear span lie in the quartic, since
\[
 \Delta(u,tv_0+w)=t^4\Delta(u,v_0+w/t)=0\quad(t\ne0).
\]
This is also impossible. Therefore the fiber is a two-plane through zero, $s=4$, and $K=\C^4\oplus V_0$ with $\dim V_0=2$. For each $v\in V_0$, the determinant vanishes identically in $u$, so every $v$ has support at most two. Irreducibility then puts $V_0$ in one coordinate two-plane; equality of dimensions makes it the whole plane. This is the row and two cells.

If neither projection contains a vector of full support, each has dimension at most three. Dimension six forces both dimensions to be three and $K$ to be their direct sum. If the missing coordinates differ, after relabeling the point $u=(1,1,1,0)$, $v=(1,1,0,1)$ lies in $K$ and has $\det B=-4$. This violates $r\leq3$. The missing coordinates therefore agree, and $K$ is the full $2\times3$ block.

\paragraph{Dimensions seven and eight.}
A seven dimensional subspace is a hyperplane of the space of the two rows. The generic rank of $B$ on it is four: otherwise its linear defining equation would divide the irreducible polynomial $\Delta$, which it cannot do. Every such hyperplane satisfies $r\leq4$. Dimension eight gives the space of the two rows itself. These hyperplanes form continuous families, and they stay in the list as families.

\subsection{Kernels inside a cross}
Let the cross have center $s$ and arm vectors $a,d\in\C^3$. On the inner $3\times3$ variables its Hessian is, up to the ordering of coordinates, $C(a)\otimes C(d)$. There are no other nonzero Hessian rows. The rank of $C(w)$ is zero if $w=0$, two if the support of $w$ has size one or two, and three if it has size three. The determinant formula in \eqref{eq:Cmatrix} proves the last assertion; a nonzero off-diagonal entry supplies a nonsingular $2\times2$ block in the other cases.

For a linear $K$, let $p,q$ count the arm coordinates that are not identically zero on it. Generically all those coordinates are simultaneously nonzero, so the generic Hessian rank is the product of the corresponding ranks just listed. If $k=7$, the cross is full and its Hessian rank is nine, exceeding the allowed eight. If $k=6$, rank at most six forces one arm to miss a coordinate. The resulting coordinate hyperplane has dimension six and contains $K$, so it equals $K$, and $K=K_6$. If $k=5$, neither arm can be identically zero, since that would put $K$ in a four dimensional row or column. Rank at most four then forces each arm to miss a coordinate. Their common coordinate support has dimension five, so $K$ is the cross with arms of two. Dimensions three and four reduce to a row or column, as already proved.

\subsection{The classification}
Combining the previous arguments gives the following list.
\begin{theorem}\label{thm:kernellist}
Up to the stated symmetries, every admissible kernel is exactly one of the following:
\begin{enumerate}
\item a three dimensional hyperplane in a row;
\item a full row;
\item a five dimensional hyperplane in a $2\times3$ block;
\item the full $2\times3$ block;
\item the row and two cells;
\item the cross with arms of two;
\item the cross $K_6$;
\item a seven dimensional hyperplane in the space of two rows;
\item the space of two rows.
\end{enumerate}
Every space on this list is admissible. The next three sections show that none of them is the endpoint kernel of a representation of size $14$.
\end{theorem}
As sets, the parameter spaces $X_k$ consist of eight copies of $\PP^3$ for $k=3$; eight points for $k=4$; forty-eight copies of $\PP^5$ and $144$ isolated points for $k=5$; $288$ points for $k=6$; twelve copies of $\PP^7$ for $k=7$; and twelve points for $k=8$. Distinct containers meet in dimensions too small to contain a common $k$-plane, so these components do not overlap.

The hyperplane families have continuous moduli. A hyperplane in a $2\times r$ block has a nonzero annihilator matrix with two rows and $r$ columns. After column scalings, each column has one of the forms $0$, $(1,0)^T$, $(0,1)^T$, or $(1,t_j)^T$ with $t_j\ne0$. Record the counts of the first three kinds and the unordered list of the $t_j$. Row scaling multiplies all $t_j$ by one common nonzero scalar, and swapping the rows inverts them and exchanges the counts of the second and third kinds. Up to these operations the $t_j$ are invariants of the hyperplane.

\section{Row hyperplanes and the smaller kernels}
\label{sec:small}
It remains to show that no representation of size $14$ has an admissible endpoint kernel. We go container by container and leave the full row to the end, because there $\rho_K=0$ and Proposition~\ref{prop:endpoint} carries no information beyond $m\geq13$. This section excludes the hyperplanes of a row and the kernels of dimensions five and six. Throughout, $E=U^\perp/U$ and the maps $R_z$ are those of Section~\ref{sec:endpoint}. A change of basis in $E$ is a change of coordinates in an auxiliary quadratic space, and not a row or column operation on the representation.

\subsection{Stabilizers of the contracted cubics}\label{sec:cubicstabilizers}
For a polynomial $P$ and a constant linear operator $G$, write $\delta_GP(y)=dP_y(Gy)$. We need two stabilizer calculations. The archives check them with complete rational coefficient matrices; here we argue directly with coefficients.

First, for any contraction $F_z(a,b,c)$ with $z$ of support at least three, the linear matrix $B(z,a)$ is generically invertible. This follows from the two nonzero evaluations of $\Delta$ in Section~\ref{sec:insidetworows}, since $B(u,v)=B(v,u)$, transported by column scaling. Consider a linear vector field on the direct sum of the three lower rows. An off-diagonal block, for example a map $P:a\mapsto c$, contributes the polynomial
\[
 b^TB(z,a)Pa,
\]
which has multidegree $(2,1,0)$ in $(a,b,c)$, and no other block contributes to that multidegree. If the vector field stabilizes $F_z$, this expression is zero. Generic invertibility of $B(z,a)$ gives $Pa=0$ generically and hence identically, so $P=0$. The other five off-diagonal blocks vanish in the same way, with the rows permuted. Thus every stabilizer is block diagonal with respect to the three rows.

\begin{lemma}\label{lem:fullstab}
For $z=\mathbf1$, every infinitesimal stabilizer of $F_z$ is
\[
 (a,b,c)\longmapsto(\alpha a,\beta b,\gamma c),
 \qquad \alpha+\beta+\gamma=0.
\]
The same description holds for every $z$ of full support, after the simultaneous diagonal column change.
\end{lemma}
\begin{proof}
Let the three row blocks be $P,Q,R$. In $\delta F$, the coefficient of $a_i b_i c_k$, for $k\ne i$, is
\[
 \sum_{l\notin\{i,k\}}(P_{li}+Q_{li}).
\]
For fixed $i$, the three choices of $k$ give the three pair sums of the three entries $P_{li}+Q_{li}$ with $l\ne i$. All pair sums zero imply that each entry is zero, since $2\ne0$. Repeating the argument with the repeated coordinate in the other two pairs of rows gives $P_{li}+R_{li}=Q_{li}+R_{li}=0$. These three equations force all off-diagonal entries of $P,Q,R$ to be zero.

For pairwise distinct $i,j,k$, the coefficient of $a_i b_j c_k$ is then $P_{ii}+Q_{jj}+R_{kk}$. Holding $i,j$ fixed and varying between the two available $k$ proves equality of the corresponding two diagonal entries of $R$. Every pair of indices can be the available pair, so $R$ is scalar. The same reasoning makes $P$ and $Q$ scalar, and the displayed coefficient makes their three scalars sum to zero. Conversely these row scalings stabilize any polynomial of multidegree $(1,1,1)$.

For general $z$ of full support, substitute $Y=\widetilde Y\diag(z)$. Every permanent term acquires $\prod_jz_j$, so
\[
 F_z(\widetilde Y\diag(z))=\Bigl(\prod_jz_j\Bigr)F_{\mathbf1}(\widetilde Y).
\]
This transports the calculation without mixing the rows.
\end{proof}

\begin{lemma}\label{lem:threestab}
For $z=(z_1,z_2,z_3,0)$ with $z_1z_2z_3\ne0$, write the lower rows as $(v_i,t_i)$, $v_i\in\C^3$. The full infinitesimal stabilizer is
\[
 v_i\longmapsto(H+\alpha_iI_3)v_i,
 \qquad t_i\longmapsto-(\alpha_j+\alpha_k)t_i,
 \qquad H^TC(z)+C(z)H=0,
\]
where $\{i,j,k\}=\{1,2,3\}$.
\end{lemma}
\begin{proof}
The cubic is
\[
 t_1v_2^TC(z)v_3+t_2v_1^TC(z)v_3+t_3v_1^TC(z)v_2.
\]
The multidegree argument above has already removed every block between different rows. In row $i$ write a general block as
\[
 v_i\longmapsto H_iv_i+p_it_i,
 \qquad t_i\longmapsto q_i^Tv_i+\lambda_it_i.
\]
A constant change of the $v$ coordinates makes the nonsingular symmetric matrix $C(z)$ the identity for this calculation. The terms of degree zero in the $t_i$ are
\[
 (q_1^Tv_1)(v_2^Tv_3)+(q_2^Tv_2)(v_1^Tv_3)
 +(q_3^Tv_3)(v_1^Tv_2).
\]
Set $v_1=e_r$ and $v_2=v_3=e_s$ with $r\ne s$. This gives $(q_1)_r=0$; all coordinates and all three rows are covered by varying $r,s$ and permuting the rows. Hence $q_i=0$.

The coefficient of $t_it_jv_k$ is the linear form $(p_i+p_j)^TC(z)v_k$. Nonsingularity gives the three equations $p_i+p_j=0$, hence every $p_i=0$. The remaining equations are
\[
 H_2^TC+CH_3+\lambda_1C=0,\quad
 H_1^TC+CH_3+\lambda_2C=0,\quad
 H_1^TC+CH_2+\lambda_3C=0.
\]
Subtracting pairs shows that the differences $H_i-H_j$ are scalar matrices. Choose scalars $\alpha_i$ satisfying $\lambda_i=-(\alpha_j+\alpha_k)$; the unique solution is obtained by adding and subtracting the three equations and dividing by two. Then $H_i-\alpha_iI$ is one common matrix $H$, and any remaining equation is $H^TC+CH=0$. Conversely the displayed family makes all coefficients vanish, so the description is complete.
\end{proof}
The twelve partial derivatives of either contracted cubic in these two cases are linearly independent. Derivatives in different rows have different multidegrees. Within one row, the four bilinear coefficients are independent because the associated matrix $B$ is generically invertible. Thus the cubic depends essentially on all twelve variables of the lower rows. As a check, the coefficient maps on all $144$ entries of a linear vector field have ranks $142$ and $138$, in agreement with the stabilizers of dimensions two and six found above.

\subsection{Three computations with $C(d)$}\label{sec:relativeC}
We record three small calculations that are used twice. First, if a constant $3\times3$ matrix $T$ satisfies
\[
 T^TC(d)+C(d)T=0\quad\text{for every }d,
\]
then $T=0$. The diagonal equations force its off-diagonal entries to be zero: for example $d_3T_{21}+d_2T_{31}=0$ forces $T_{21}=T_{31}=0$. The off-diagonal equations then make every pair of diagonal entries sum to zero, hence all diagonal entries vanish.

Second, suppose $H(d)$ has homogeneous linear entries and $H(d)^TC(d)+C(d)H(d)=0$. The diagonal equations give
\[
 H_{21}=a d_2,\ H_{31}=-a d_3,\quad
 H_{12}=b d_1,\ H_{32}=-b d_3,\quad
 H_{13}=c d_1,\ H_{23}=-c d_2
\]
for constants $a,b,c$. For instance $d_3H_{21}+d_2H_{31}=0$ and coprimality of $d_2,d_3$ give the first pair, with a constant quotient because the entries are linear. The three off-diagonal equations give
\[
 H_{11}+H_{22}=ad_1+bd_2,\quad
 H_{11}+H_{33}=cd_3-ad_1,\quad
 H_{22}+H_{33}=-cd_3-bd_2.
\]
Solving them yields the complete matrix
\begin{equation}\label{eq:relativeH}
 H(d)=
 \begin{pmatrix}
 bd_2+cd_3&bd_1&cd_1\\
 ad_2&ad_1-cd_3&-cd_2\\
 -ad_3&-bd_3&-ad_1-bd_2
 \end{pmatrix}
 =K_0C(d),
\end{equation}
where $K_0$ is the constant skew matrix with upper entries $(c,b,a)$. In particular the coefficients of $d_1^2,d_2^2,d_3^2$ in $\tr(H(d)^2)$ are $2a^2,2b^2,2c^2$. A pencil of this form that is nilpotent for every $d$ must be zero.

Third, if constant matrices $H,V$ satisfy $C(d)H=VC(d)$ for every $d$, then $H=V=\sigma I$. The coefficient matrix of $d_1$ exchanges the second and third basis vectors and kills the first. Comparing its first column and first row forces $H_{21}=H_{31}=V_{12}=V_{13}=0$. Repeating with $d_2,d_3$ makes $H,V$ diagonal. The nonzero off-diagonal entries of the three $C(d)$ coefficients give $H_{jj}=V_{ii}$ for every $i\ne j$, so all six diagonal values are equal. In archive~A these three calculations appear as coefficient matrices of ranks $9$, $24$ and $17$, on $9$, $27$ and $18$ unknowns.

\subsection{Hyperplanes in a row}\label{sec:rowthree}
Let $K$ be any hyperplane in the first row. Then $a=13=h$, so $U^\perp=U$. For every $z\in K$, $M_zU\subseteq U$, and it induces a nilpotent linear operator $S_z$ on the thirteen endpoint coordinates. Write those coordinates as $(Y,t)$, with $Y$ the twelve entries of the lower rows and $t$ one complementary coordinate in the first row. The family $S_z$ is linear in $z$.

The part of $f(sz+Y+t)$ of degree three in $(Y,t)$ is $sF_z(Y)$. The Hessian at a point of a row is zero. Lemma~\ref{lem:cubicresolvent} therefore gives
\begin{equation}\label{eq:rowthreeident}
 \delta_{S_z}F_z=0,
 \qquad F_z(Y)=-b(Y,t)D(S_z(Y,t))c(Y,t).
\end{equation}
For generic $z\in K$, its support has size at least three and $F_z$ is essential in all twelve $Y$ variables. The coefficient of $t$ in $\delta_{S_z}F_z$ is a constant linear combination of those twelve derivatives. Their independence forces the $t$-to-$Y$ block of $S_z$ to be zero. Thus the induced $Y$ block is nilpotent, and the scalar $t$-to-$t$ block is zero as well.

If $K$ is not a coordinate hyperplane, its generic vector has full support. Lemma~\ref{lem:fullstab} makes the $Y$ block a scalar on each lower row; nilpotence makes all three scalars zero. Since $S_z$ is polynomial in $z$, that block vanishes for every $z\in K$.

If $K$ is a coordinate hyperplane, take $z_4=0$. Lemma~\ref{lem:threestab} applies on the dense subset where $z_1z_2z_3\neq0$. The one dimensional $t_i$ blocks have eigenvalues $-(\alpha_j+\alpha_k)$, so nilpotence forces all $\alpha_i=0$. The three $v_i$ blocks are a common matrix $H_z$ satisfying $H_z^TC(z)+C(z)H_z=0$. Because $S_z$ is linear in $z$, $H_z$ is linear as well. Formula \eqref{eq:relativeH} and the three square coefficients in its trace show that $H_z=0$. Polynomiality again extends this to all $z$.

In either case $S_z(Y,t)=\ell_z(Y,t)e_t$. The second identity in \eqref{eq:rowthreeident} becomes
\[
 F_z(Y)=-\ell_z(Y,t)b(Y,t)D(e_t)c(Y,t).
\]
For nonzero $z$, the left side is a nonzero irreducible cubic by Lemma~\ref{lem:cubicirreducible}. The right side is either zero or a product of a homogeneous linear form and a homogeneous quadratic. Neither is possible. This excludes every hyperplane in a row, whatever the support of its annihilator, and transposition excludes the hyperplanes in a column.

\subsection{Syzygies among $2\times2$ subpermanents}\label{sec:syzygies}
Let
\[
 P_1=u_2v_3+u_3v_2,\quad P_2=u_1v_3+u_3v_1,\quad
 P_3=u_1v_2+u_2v_1.
\]
We use two properties:
\begin{equation}\label{eq:syzygyproperties}
 \sum_i\ell_iP_i=0,\ \deg\ell_i=1\ \Longrightarrow\ \ell_i=0;
 \qquad
 \sum_iq_iP_i=0,\ \deg q_i=2\ \Longrightarrow\ q=JP
\end{equation}
for a constant skew $3\times3$ matrix $J$.

To prove them, order the six variables as $(u_1,u_2,u_3,v_1,v_2,v_3)$. Let $\mathcal M_d$ be their degree-$d$ monomials in lexicographic order. Define the integer matrix
\[
 (T_d)_{\mu,(i,\nu)}=[\mu](P_i\nu),
 \qquad \mu\in\mathcal M_{d+2},\quad i=1,2,3,\quad\nu\in\mathcal M_d.
\]
All its entries are $0$ or $1$. Rows that are identically zero may be omitted without changing its rank. Exact elimination gives
\[
 \rk T_1=18=\#\text{columns},\qquad
 \rk T_2=60=63-3.
\]
The three vectors $P_je_i-P_ie_j$, $1\leq i<j\leq3$, are independent elements of $\ker T_2$, since the $P_i$ are linearly independent. The rank therefore proves that they form its entire kernel. This establishes both statements in \eqref{eq:syzygyproperties}. The integer matrices and kernel vectors are stored in \file{k6_attempt/two_by_three_syzygy_certificate.json} in archive~A, and its checker regenerates them.

\begin{lemma}\label{lem:nolinearminor}
There is no homogeneous linear matrix $L(u,v)$ of rank at most two whose row space over $\C(u,v)$ contains $(P_1,P_2,P_3)$.
\end{lemma}
\begin{proof}
Rank zero cannot contain a nonzero vector. If the rank is one, choose a nonzero linear row of $L$. It has a nonzero syzygy with linear coefficients: for two entries use their signed interchange, or, if only one entry is nonzero, put any nonzero linear form in a zero coordinate. Every row of $L$ is rationally proportional to the chosen row, so that same polynomial vector annihilates every row and also $P$. This contradicts the first statement of \eqref{eq:syzygyproperties}.

If the rank is two, choose two independent rows and take their cross product $q$. It is a nonzero homogeneous quadratic vector annihilating every row and hence $P$. The second statement gives $q=JP$ with $J\ne0$ skew. Such a $3\times3$ matrix has rank two. For each row $\ell$ of $L$, the equation $\ell JP=0$ is a syzygy of $P$ with linear coefficients. The first statement forces $\ell J=0$. All rows therefore belong to the one dimensional left kernel of $J$, contradicting rank two.
\end{proof}

\subsection{The full $2\times3$ block}\label{sec:block2x3}
Let the kernel rows be $(u_1,u_2,u_3,0)$ and $(v_1,v_2,v_3,0)$, and let $w,t$ be the other two rows. There are also two endpoint variables in column four of the kernel rows. The quotient has dimension six. The part of $f$ of degree two in the endpoint variables is
\[
 w_4(P_1t_1+P_2t_2+P_3t_3)+t_4(P_1w_1+P_2w_2+P_3w_3).
\]
For $i=1,2,3$, put $L_i=R_z(w_i)$ and $N_i=R_z(t_i)$. The six vectors are mutually orthogonal, because this quadratic part has no term involving two of $w_1,w_2,w_3,t_1,t_2,t_3$. Their span $T$ is isotropic in a nondegenerate six dimensional space, so $\dim T\leq3$.

The functionals on $T$ given by pairing with $R_z(w_4)$ and $R_z(t_4)$ are independent over $\C(u,v)$. On the $L_i$ their values are respectively $0,P_i$, and on the $N_i$ they are $P_i,0$. Since $P\ne0$, a linear relation between the functionals has both coefficients zero. The span of the $L_i$, which is killed by the first functional, consequently has dimension at most two. But the matrix whose columns are $L_i$ is homogeneous linear in $(u,v)$, and pairing its columns with $R_z(t_4)$ puts $P$ in its row space. This contradicts Lemma~\ref{lem:nolinearminor}. The argument does not use the equality case of the endpoint inequality: here the Hessian has rank four and $E$ has dimension six.

\subsection{Hyperplanes of the $2\times3$ block}\label{sec:block2x3hyp}
Now let $K$ be a five dimensional hyperplane in the same block. The quadratic part of $f$ in the endpoint variables has the same shape as before, with $P$ replaced by its restriction $P|_K$. The entries of $P|_K$ have greatest common divisor one. Indeed their common zero set in the ambient six dimensional block has dimension three by Lemma~\ref{lem:zero2}: its components are row spaces of dimension three, column spaces of dimension two, and $2\times2$ quadric cones of dimension three. A common nonconstant factor on the five dimensional linear space $K$ would produce a four dimensional hypersurface of common zeros, which cannot fit into that three dimensional set.

The block of the Hessian between the $w$ and the $t$ variables has generic rank two, since $P|_K$ is not identically zero, so the Hessian has rank four and fills the four dimensional space $E$. By~\eqref{eq:tightkernel}, every row of the quotient map lies in the row space of the Hessian. Restricted to the $w$ variables, its first three entries are therefore a rational multiple of $P|_K$. If a rational scalar times a primitive polynomial vector is polynomial, that scalar is polynomial: write it as a reduced fraction and observe that its denominator must divide every vector entry. Since the entries of $P|_K$ have degree two, no nonzero such multiple has degree one. The first three entries must vanish. Every row of the restriction is therefore supported on the fourth coordinate, so the restriction has rank at most one, while its pairing with the restriction to the $t$ variables has rank two. This excludes every hyperplane of the block.

\subsection{The cross with arms of two}\label{sec:arms2}
Take kernel variables $s=x_{11}$, $a=x_{12}$, $b=x_{13}$, $d_1=x_{21}$, $d_2=x_{31}$. Denote the three inner columns by $p,q,r$. The other endpoint coordinates $x_{14},x_{41}$ lie in the radical of the Hessian. The endpoint quadratic is
\[
 (bp+aq)^TC(d_1,d_2,0)r.
\]
Its Hessian rank is four, filling $E$. The restriction of $R_z$ to $(p,q)$ must have rank at least two, since its pairing with the $r$ columns has rank two.

Consider one linear row $(u^T,v^T)$ of this restriction. It annihilates every radical vector $(aw,-bw)$, so $au=bv$ as polynomial vectors. Coprimality of $a,b$ and linearity imply $u=bu_0$, $v=au_0$ for one constant vector $u_0$. It also annihilates the kernel vector $(d_1,-d_2,0)$ of $C(d_1,d_2,0)$, placed in either $p$ or $q$. Hence $d_1(u_0)_1-d_2(u_0)_2=0$ identically, which gives $(u_0)_1=(u_0)_2=0$. Every row is a constant multiple of $(be_3^T,ae_3^T)$. The restriction thus has rank at most one, which contradicts the rank two of its pairing with the $r$ columns.

\subsection{The row and two cells}\label{sec:row2cells}
Let the kernel rows be $u=(u_1,u_2,u_3,u_4)$ and $(a,b,0,0)$. The block of the Hessian between the two endpoint rows is
\[
 B=
 \begin{pmatrix}
 0&0&bu_4&bu_3\\
 0&0&au_4&au_3\\
 bu_4&au_4&0&bu_1+au_2\\
 bu_3&au_3&bu_1+au_2&0
 \end{pmatrix}.
\]
Its generic rank is three and its kernel is spanned by $(a,-b,0,0)$. Thus the Hessian has rank six and fills the six dimensional space $E$. A homogeneous linear row annihilating that kernel has first entries $(b\lambda,a\lambda)$ with constant $\lambda$. Consequently the quotient columns on $w_1,w_2$ are $bL,aL$ for one constant vector $L$, and those on $t_1,t_2$ are $bN,aN$ for another constant vector $N$.

Since the Hessian has no terms with both variables in the same endpoint row, $L$ and $N$ are isotropic, and the zero upper left block of $B$ makes them orthogonal. They are independent: if $N$ were proportional to $L$, the pairing $\langle R_z(w_3),L\rangle=0$ within the $w$ row would imply $\langle R_z(w_3),N\rangle=0$, whereas the entries of $B$ require that last pairing to be $u_4$. Thus $T=\Span(L,N)$ is a constant isotropic two-plane.

Specialize $u_3=u_4=0$. All quotient columns are then orthogonal to both $L$ and $N$, because the relevant entries of $B$ are multiples of $u_3$ and $u_4$, and the pairings within one row vanish identically. Their Gram matrix factors through $T^\perp/T$, a nondegenerate space of dimension $6-2\cdot2=2$, so it has rank at most two. But after the specialization $B$ keeps its lower $2\times2$ block, with off-diagonal entry $bu_1+au_2$, which has rank two generically. So the specialized Hessian has rank at least four, a contradiction.

\subsection{The kernel $K_6$}\label{sec:K6}
For $K_6$ at $m=14$ the inequality is an equality: $h-a=3=\rho_K/2$. Equality forces the classes $M_z\Phi(y)+U$ to fill $E$, which fixes the action of $D(z)$ on $U^\perp$ up to a Witt transformation. Comparing the cubic part of $bD^2c=f$ then forces the remaining part of $D$ into a block form governed by one matrix $G$ and one linear form $\ell$. The coefficient of $x_{11}$ in $f$ is the permanent $\per_3$ of rows and columns $2,3,4$. Viewing $\per_3$ as a trilinear function of $p,q,r$, the resulting identity reads $\per_3(Gp,q,r)+\per_3(p,Gq,r)=0$, and the linear operator $G\mapsto\per_3(Gp,q,r)+\per_3(p,Gq,r)$ is injective. So $G=0$, and then $\per_3$ is a linear form times a quadric. This contradicts the irreducibility of $\per_3$. The details follow.

Write the kernel coordinates as
\[
 s=x_{11},\quad a=x_{12},\quad b=x_{13},\quad
 d=(x_{21},x_{31},x_{41}).
\]
Write the ten endpoint coordinates as $y=(p,q,r,t)$, where $p,q,r\in\C^3$ are the lower entries of columns $2,3,4$, and $t=x_{14}$. The endpoint quadratic is
\begin{equation}\label{eq:K6quad}
 (bp+aq)^TC(d)r.
\end{equation}
Its Hessian has rank six and radical $r=0$, $bp+aq=0$, with $t$ arbitrary. The quotient $E$ has dimension six, so by~\eqref{eq:tightkernel} $R_z$ has exactly this kernel generically.

In particular $R_z(t)=0$. The identity $R_z(aw,-bw,0,0)=0$ and linearity in the independent kernel variables show that
\[
 R_z(p,q,0,0)=L(bp+aq)
\]
for one constant map $L:\C^3\to E$. The pairing with $r$ has rank three, so $L$ is injective. Its image is isotropic because~\eqref{eq:K6quad} has no terms of degree two in $(p,q)$, and so it is Lagrangian in $E$. Choose a fixed Witt decomposition of $E$ in which $L(v)=(v,0)$. The required pairing with $r$ forces
\[
 R_z(p,q,r,t)=(bp+aq+T(z)r,C(d)r),
 \qquad T(z)^TC(d)+C(d)T(z)=0.
\]
The coefficients of $s,a,b$ in $T$ vanish by the first calculation in Section~\ref{sec:relativeC}. The second gives $T(d)=K_0C(d)$ for a constant skew $K_0$. The constant isometry $(v,w)\mapsto(v-K_0w,w)$ removes it. We obtain the quotient formula
\begin{equation}\label{eq:K6R}
 R_z(p,q,r,t)=(bp+aq,C(d)r).
\end{equation}
In particular $R_s=0$, so $M_sU\subseteq U$. Let $A$ be its induced operator on the endpoint coordinates; it is nilpotent. Self-adjointness also gives an induced self-adjoint operator $E_s$ on $E$.

Take the coefficient of $sz^2$ in the third moment $\langle U,M_{sz+z'}^3U\rangle=0$, with the notation arranged so $s$ denotes the center direction and $z$ the other kernel variables. The three placements of $M_s$ give
\begin{equation}\label{eq:K6M3}
 A^TQ_z+Q_zA+R_z^T J E_sR_z=0,
 \qquad Q_z=R_z^TJR_z.
\end{equation}
If $v\in\ker Q_z=\ker R_z$, this identity gives $Q_zAv=0$. Hence $A$ preserves every radical of \eqref{eq:K6quad}. Their intersection is the $t$-line and their span is the $(p,q,t)$ space. On its quotient by the $t$-line, preservation of all subspaces $bp+aq=0$ forces the same matrix on $p$ and $q$ and no off-diagonal block: set $(a,b)=(1,0),(0,1),(1,1)$ in turn. Thus
\[
 A(p,q,r,t)=
 (Gp+Pr,\ Gq+Qr,\ Hr,\ \alpha p+\beta q+\gamma r+\kappa t).
\]
Write the self-adjoint operator on the hyperbolic quotient as
\[
 E_s=\begin{pmatrix}X&Y\\Z&X^T\end{pmatrix},\qquad Y=Y^T,\quad Z=Z^T.
\]
Substitute \eqref{eq:K6R} in \eqref{eq:K6M3}. Put $w=bp+aq$ for this calculation. The terms quadratic in $w$ give $Z=0$. The mixed $w,r$ terms give
\[
 C(d)H+(G^T+X^T)C(d)=0.
\]
The terms quadratic in $r$ give
\[
 C(d)(bP+aQ)+(bP+aQ)^TC(d)+C(d)YC(d)=0.
\]
Independence of $a,b$ gives $C(d)P+P^TC(d)=0$ and the same equation for $Q$. The first calculation of Section~\ref{sec:relativeC} yields $P=Q=0$. Generic invertibility of $C(d)$ then gives $Y=0$. The third calculation of that section, applied with $V=-G^T-X^T$, gives $H=V=\sigma I$. Nilpotence of $A$ gives $\sigma=\kappa=0$, and $G$ is nilpotent. Therefore
\begin{equation}\label{eq:K6A}
 A(p,q,r,t)=(Gp,Gq,0,\ell(y)),\qquad
 E_s=\diag(-G,-G^T).
\end{equation}

Set the other five kernel variables to zero. The part of $f$ of degree three in $y$ becomes $sP(y)$ with $P(y)=\per_3(p,q,r)$. The center direction has zero Hessian, so Lemma~\ref{lem:cubicresolvent} gives $\delta_AP=0$ and $P=-bD(Ay)c$. Using \eqref{eq:K6A}, the first equation is
\[
 \per_3(Gp,q,r)+\per_3(p,Gq,r)=0.
\]
For distinct $i,j,k$, its coefficient of $p_iq_ir_k$ is $2G_{ji}$, so every off-diagonal entry of $G$ vanishes. Its coefficient of $p_iq_jr_k$ is $G_{ii}+G_{jj}$, so all diagonal pair sums vanish and $G=0$. Now $Ay=\ell(y)e_t$ and
\[
 \per_3(p,q,r)=-\ell(y)b(y)D(e_t)c(y),
\]
contradicting Lemma~\ref{lem:cubicirreducible}. This excludes $K_6$.

The kernels of Theorem~\ref{thm:kernellist} that remain are the full row, the hyperplanes of two rows, and two rows. The next section treats the last two.

\section{Kernels in two rows}
\label{sec:tworows}
This section excludes the space of two rows and every seven dimensional hyperplane in it. Here $b$ and $c$ may each involve both of the other two rows, so the endpoints can be mixed. The proof of Proposition~\ref{prop:endpoint} still factors the Hessian through the ten dimensional space $E$. We convert this into a linear factorization of $B$ through at most five channels. The conversion uses an operator $T_F$ on the even part of the exterior algebra of $\C^5$ for each isotropic frame $F$. These operators satisfy $\tr(T_FT_G)=\det(F^TJG)$ and $\rk T_F=1$, where $J$ is the Gram matrix of the form on $E$. Theorem~\ref{thm:fivechannels} then applies. For a hyperplane, $E$ has dimension eight, and the same frames give a condition on the factors of $\det B$ restricted to the hyperplane.

\subsection{No factorization through five channels}\label{sec:fivechannels}
\begin{theorem}\label{thm:fivechannels}
There are no homogeneous linear matrices $P(u,v)$ of size $4\times5$ and $Q(u,v)$ of size $5\times4$ satisfying $B(u,v)=P(u,v)Q(u,v)$.
\end{theorem}
\begin{proof}
Write $P=P_u+P_v$ and $Q=Q_u+Q_v$, separating dependence on the two sets of variables. The pure and mixed bidegrees give
\begin{equation}\label{eq:puremixed}
 P_uQ_u=P_vQ_v=0,\qquad B=P_uQ_v+P_vQ_u.
\end{equation}
We first eliminate small pure ranks, and then reduce the remaining factorizations to the ten directions of rank two in Lemma~\ref{lem:tenranktwo}.

Let $T_{ijab}=[u_av_b]B_{ij}$. It is one when $i,j,a,b$ are pairwise distinct and zero otherwise. Flatten this tensor with row index $(i,b)$ and column index $(a,j)$. Viewed as a map on $4\times4$ matrices, its image is
\[
 \mathcal S=\{Z:Z=Z^T,\ Z_{ii}=0\},\qquad \dim\mathcal S=6.
\]
Indeed an off-diagonal output is the sum of the two input entries on the complementary pair of indices; all six such sums vary independently. An ordered product $V(v)U(u)$ through $r$ channels is a sum of $r$ tensors of rank one in this flattening, so its flattened rank is at most $r$.

Projecting the first output coordinate modulo a fixed nonzero line preserves the rank six. An element of the kernel of that projection would be a symmetric matrix with image in a line, hence of rank at most one. A symmetric matrix of rank one with zero diagonal is zero. More generally, for a fixed $t$-plane $L\subseteq\C^4$,
\begin{equation}\label{eq:Sintersection}
 \dim\{Z\in\mathcal S:\im Z\subseteq L\}\leq t(t-1)/2.
\end{equation}
Choose a matrix $R$ of full column rank with image $L$. Every symmetric matrix whose image is in $L$ is uniquely $RHR^T$ with $H$ symmetric. Choose $t$ independent rows of $R$. Their squares give $t$ independent linear conditions on $H$ when the corresponding diagonal entries are set to zero: a change of basis makes those rows coordinate covectors. Subtracting these conditions from $t(t+1)/2$ proves \eqref{eq:Sintersection}.

There cannot be four ordered channels and one reversed channel:
\begin{equation}\label{eq:fourplusone}
 B=V(v)U(u)+(Au)(v^TD).
\end{equation}
In the chosen flattening, the reversed term is the map $K:M\mapsto AMD^T$. If $p=\rk A$ and $q=\rk D$, its image has dimension $pq$. A symmetric matrix in its intersection with $\mathcal S$ has image in both $\im A$ and $\im D$, so \eqref{eq:Sintersection} bounds that intersection by $\binom{\min(p,q)}2$. Projecting $T-K$ modulo $\im K$, and separately modulo $\mathcal S$, gives
\[
 \rk(T-K)\geq\max(6,pq)-\dim(\mathcal S\cap\im K)
 \geq\max(6,pq)-\binom{\min(p,q)}2>4.
\]
For $\min(p,q)\leq2$ the bound is at least five. For $(p,q)=(3,3),(3,4),(4,4)$ it is six, nine, ten, respectively; the transposed cases are identical. This contradicts the bound of four channels for the ordered part.

None of $P_u,Q_u,P_v,Q_v$ can have rank zero or one. Rank zero leaves an ordered factorization through five channels, contradicting rank six. For rank one we use an elementary fact: a linear space of matrices of rank at most one has a common image line or a common row line. Fix a nonzero element $ab^T$. Every other element of rank one must share its image line or row line, since otherwise their sum has rank two. If elements of both kinds occur without both lines in common, their sum also has rank two. Thus one kind holds throughout.

Apply this fact to $P_u$. A common image line can be projected away in \eqref{eq:puremixed}; the $P_uQ_v$ term disappears, leaving an ordered factorization through five channels of a matrix whose flattened rank is still six. For a common row line, write $P_u=\alpha(u)\beta^T$ with $\beta$ constant. The pure identity gives $\beta^TQ_u=0$. A constant internal basis makes $\beta$ the fifth coordinate covector. The $P_vQ_u$ term then has only four channels and the other term is one reversed channel, forbidden by \eqref{eq:fourplusone}. Transposition exchanges $P,Q$, and exchange of $u,v$ handles both other pure factors. All four pure ranks are therefore at least two.

Put $p=\rk P_u$ and $q=\rk Q_u$ over $\C(u)$. The pure identity gives $p+q\leq5$, while the generic rank four of $B$ gives $p+q\geq4$. If $p+q=4$, bases adapted to $\im P_u$ and $\ker Q_u$ put the mixed matrix into the block form
\[
 \begin{pmatrix}*_{p\times q}&X_{p\times p}\\
 Y_{q\times q}&0_{q\times p}\end{pmatrix}.
\]
Its determinant is $\pm\det X\det Y$. The square blocks are linear in $v$, and both determinants are nonzero because $\det B\ne0$. Their positive degrees contradict irreducibility over $\C(u)[v]$ from Lemma~\ref{lem:crossdet}. Thus $(p,q)$ is $(2,3)$ or $(3,2)$; transposing lets us assume $(2,3)$.

Choose a nonzero rational vector $k(u)$ spanning $\ker Q_u$. Then
\[
 B(u,v)k=P_uQ_vk.
\]
The coefficient tensor is symmetric in all four indices, so $B(u,v)k=B(u,k)v$. It follows that $\rk B(u,k)\leq2$. For $u$ of full support, put $z_i=k_i/u_i$. The diagonal congruence in Lemma~\ref{lem:crossdet} and Lemma~\ref{lem:tenranktwo} put $z$ in one of ten fixed lines. The rational image of an irreducible variety in a finite union of closed lines lies in one of them. Hence $k(u)=\lambda(u)Zu$ for a fixed diagonal $Z$ from that list. The mixed identity gives
\[
 \im B(u,Zu)\subseteq\im P_u,
\]
and both sides have generic dimension two, so they are equal.

For a star line, take $Z=\diag(-1,1,1,1)$. Substitution gives
\[
 \im B(u,Zu)=\Span\left(e_1,(0,u_3u_4,u_2u_4,u_2u_3)^T\right).
\]
The last three entries of each linear column of $P_u$ are a rational multiple of a primitive quadratic vector. Primitivity makes that multiplier polynomial, and degree makes the multiple zero. All columns lie in the $e_1$ line, contradicting rank two. Permutations exclude every star.

For a pair difference, take $Z=\diag(1,-1,0,0)$. Now $\im B(u,Zu)$ is
\[
 \Span(p(u),q(u)),\qquad p=(-u_2,u_1,0,0)^T,\quad q=(0,0,u_4,u_3)^T.
\]
Every linear polynomial section of this plane is a constant combination of $p,q$: apply the divisibility argument for primitive vectors to each of the two coordinate pairs. Therefore $P_u=p\alpha^T+q\beta^T$ with constant rows $\alpha,\beta$. The identity $Q_u(u)(u_1,-u_2,0,0)^T=0$ gives
\[
 (Q_u)_{\mathrm{col}\,1}=u_2\eta,\qquad
 (Q_u)_{\mathrm{col}\,2}=u_1\eta
\]
with constant $\eta$, by coprimality and linearity. Entry $(2,1)$ of the mixed identity is
\[
 B_{21}=u_1\alpha^T(Q_v)_{\mathrm{col}\,1}
       +u_2(P_v)_{\mathrm{row}\,2}\eta.
\]
Its right side belongs to $(u_1,u_2)$, whereas $B_{21}=u_3v_4+u_4v_3$ does not. Setting $u_1=u_2=0$ gives a contradiction. Permutations exclude all six pair differences, and this exhausts the possible factorizations.
\end{proof}

\subsection{Two isotropic frames}
Let $K$ be the space of the two middle rows $(u,v)$, and call the other two rows $(a,c)$. The endpoint image has dimension eight, so $E$ has dimension ten. Splitting the quotient map according to the rows $a$ and $c$ gives linear $10\times4$ frames $A,C$ satisfying
\begin{equation}\label{eq:ACgram}
 A^TJA=C^TJC=0,\qquad A^TJC=B(u,v).
\end{equation}
Each frame has generic rank four because $B$ is invertible generically. We now derive from~\eqref{eq:ACgram} a factorization of $B$ through five channels, which Theorem~\ref{thm:fivechannels} forbids.

\subsection{The Clifford trace identity}\label{sec:clifford}
Let $V$ be five dimensional and identify the hyperbolic space of dimension ten with $V\oplus V^*$. On $\Lambda^*V$ define
\[
 c(v+\phi)(\omega)=v\wedge\omega+\iota_\phi\omega.
\]
The contraction rule $\iota_\phi(v\wedge\omega)=\phi(v)\omega-v\wedge\iota_\phi\omega$ gives
\[
 c(x)c(y)+c(y)c(x)=J(x,y)I.
\]
Even products preserve $S^+=\Lambda^{\mathrm{even}}V$, of dimension $1+10+5=16$. For an isotropic four-frame $F=(f_1,f_2,f_3,f_4)$ set
\[
 T_F=c(f_1)c(f_2)c(f_3)c(f_4)|_{S^+}.
\]
For two such frames,
\begin{equation}\label{eq:cliffordtrace}
 \tr(T_FT_G)=\det(F^TJG).
\end{equation}
To prove the normalization, take dual hyperbolic bases $e_i,f_i$. The ten vectors $e_i+f_i,e_i-f_i$ are orthogonal with squared norms $+2,-2$, so their Clifford operators square to $+I,-I$. For an ordered subset $I$ of four elements let $\Gamma_I$ be the product of its operators. If $I=J$, anticommuting the factors contributes $(-1)^6=1$, and
\[
 \tr_{S^+}(\Gamma_I^2)=16\prod_{i\in I}\epsilon_i,
 \qquad\epsilon_i\in\{1,-1\}.
\]
This is the Gram determinant on $\Lambda^4$, since $2^4=16$. If $I\ne J$, the product is a scalar times a nonempty even word using at most eight of the ten basis indices. Choose one index inside and one outside the word. Their even Clifford product preserves $S^+$ and conjugates the word to its negative; hence its trace is zero. Bilinearity gives the pairing on all of $\Lambda^4$. An isotropic frame has no internal contractions, so its Clifford product equals the Clifford image of its wedge product. This proves \eqref{eq:cliffordtrace}.

For an independent isotropic frame, choose isotropic duals $G$ with $F^TJG=I_4$. Start with any $B$ satisfying $F^TJB=I$, put $H=B^TJB$, and replace $B$ by $B-\frac12FH$. The cross pairing stays $I$, while the pairing of $B$ with itself becomes $H-\frac12H-\frac12H=0$. Successively anticommuting each dual factor past the four isotropic factors gives
\[
 T_FT_GT_F=T_F;
\]
each cross anticommutator is its Kronecker delta and the sign for four factors is $(-1)^6=1$. Thus $T_FT_G$ is idempotent and has the same image as $T_F$. Its trace is one by \eqref{eq:cliffordtrace}, so its rank and the rank of $T_F$ are one.

When $F$ is linear in the middle variables, $T_F$ is a homogeneous quartic polynomial matrix. Any nonzero homogeneous polynomial matrix of rank one factors as
\[
 T_F=g_Fp_Fq_F^T,\qquad
 \deg g_F+\deg p_F+\deg q_F=4,
\]
with primitive homogeneous polynomial vectors $p_F,q_F$. Divide a nonzero column by its homogeneous gcd; every other column is a rational multiple of the resulting primitive vector, hence a polynomial multiple. Extract the gcd of those multiples to obtain $g_F$ and the primitive row. Homogeneity gives the degree equality.

Apply this to the frames $A,C$. The trace identity is
\[
 \det B=g_Ag_C(q_A^Tp_C)(q_C^Tp_A).
\]
The determinant is irreducible of degree eight. The two factors $g_A,g_C$ have degrees at most four, so they are nonzero constants. One of the two remaining nonzero homogeneous factors must be constant. Interchanging the frames if necessary, take $q_A^Tp_C$ constant. Its degree is the sum of nonnegative degrees, so both $q_A$ and $p_C$ are constant. Put $q=q_A\ne0$.

For every column $a_i$ of $A$, isotropy gives $T_Ac(a_i)|_{S^-}=0$, since that isotropic Clifford factor appears twice and anticommutes with the other three. This factorization of $T_A$ implies $q^Tc(a_i)|_{S^-}=0$. Define
\[
 L=\{x\in V\oplus V^*:q^Tc(x)|_{S^-}=0\}.
\]
It is a fixed isotropic subspace. Indeed for $x,y\in L$,
\[
 J(x,y)q^T=q^T(c(x)c(y)+c(y)c(x))|_{S^+}=0,
\]
so $J(x,y)=0$. Therefore $\dim L\leq5$. All columns of $A$ lie in $L$, and a constant basis matrix $L_0$ gives $A=L_0A_0$ with $A_0$ linear. Consequently
\[
 B=A_0^T(L_0^TJC)
\]
factors linearly through at most five channels, contrary to Theorem~\ref{thm:fivechannels}. This excludes the space of two rows.

The argument stops at size fourteen. In a hyperbolic space of dimension twelve the module is $\Lambda^{\mathrm{even}}\C^6$, and wedge multiplication by $e_1e_2e_3e_4$ has two independent outputs, from $1$ and $e_5e_6$. So it has rank two, and the construction of size fifteen is not excluded.

\subsection{Hyperplanes of two rows}\label{sec:K7}
Let $K$ be a hyperplane in the space of two rows. Its generic Hessian rank is eight: the polynomial $\Delta(u,v)=\det B(u,v)$ is irreducible of degree eight, and therefore cannot vanish identically on a hyperplane. Here $a=9$ and $E=U^\perp/U$ has dimension eight. The endpoint direction left out of $K$ in the two rows is in the common radical of the Hessians on $K$. The maps to $E$ are surjective at a generic point, so that direction maps to zero. The other two rows give isotropic four-frames $A(z),C(z)$ in $E$ satisfying
\[
 A^TJA=C^TJC=0,\qquad A^TJC=B|_K.
\]
Their cross pairing is generically invertible.

We can choose one constant Witt decomposition in which the upper $4\times4$ blocks of both frames are generically invertible. To justify this choice, specialize at a point where the two frames are independent and their cross pairing is nondegenerate. Their spans are complementary Lagrangians. Relative to these two Lagrangians, the graph of an invertible skew-symmetric $4\times4$ matrix is a Lagrangian transverse to both. Use that graph as the lower Lagrangian and choose any Lagrangian complement. The desired two determinants are then nonzero at the chosen point, and hence are nonzero polynomials.

Write the frames in these coordinates as $\binom{A}{C}$ and $\binom{B}{D}$. The rational matrices $S=CA^{-1}$ and $T=DB^{-1}$ are skew-symmetric, and their cross pairing is $A^T(T-S)B$. Thus
\begin{equation}\label{eq:oddobstruction}
 \det(B|_K)=\det A\det B\,\Pf(T-S)^2.
\end{equation}
In the unique factorization ring of polynomials on $K$, every irreducible factor occurring to odd multiplicity on the left must divide $\det A\det B$. Each of these two determinants has degree four. In particular, an irreducible factor of odd multiplicity and degree greater than four makes~\eqref{eq:oddobstruction} impossible.

\subsection{Noncoordinate hyperplanes}\label{sec:hyperplanefactors}
We now show that this obstruction excludes every noncoordinate hyperplane. Write its equation as
\[
 L=\sum_{i=1}^4\alpha_i u_i+\beta_i v_i.
\]
On the open set $u_1u_2u_3u_4\ne0$, put $z_i=v_i/u_i$ and $U=u_1u_2u_3u_4$. Then
\[
 \Delta=-4U^2q(z),\qquad q=e_1e_3-4e_4,\qquad
 L=\sum_i h_i(z)u_i,\quad h_i=\alpha_i+\beta_i z_i.
\]
The polynomial $q$ is irreducible, as proved in Section~\ref{sec:targetfacts}.

Suppose first that at least two $h_i$ are nonzero polynomials. Over the generic point of the irreducible hypersurface $q=0$, the equation $\sum_i h_i u_i=0$ cuts out an irreducible hyperplane in the four dimensional torus. It is reduced: solve for one $u_i$ with $h_i\ne0$ on a suitable open set. Thus there is exactly one irreducible component dominating $q=0$, and its multiplicity is one. A further component can occur only over a locus where all the $h_i$ vanish. Indeed, where exactly one coefficient is nonzero the torus fiber is empty; where at least two are nonzero it remains an irreducible three dimensional hyperplane section. A common zero locus of codimension at least two in $q=0$ has dimension at most one, and even a full four dimensional torus fiber over it has dimension at most five. It cannot be a component of the six dimensional intersection under consideration.

If a nonzero $h_i$ is constant, their common zero locus is empty. Otherwise each nonzero $h_i$ fixes its own coordinate $z_i$. Three fixed coordinates leave dimension at most one. With just two fixed coordinates $z_i=A,z_j=B$, a two dimensional common zero locus exists only if their entire affine coordinate plane is contained in $q=0$. The cubic terms in the two remaining variables first give $A+B=0$. Substituting $B=-A$ then gives
\[
 q=-A^2(z_k-z_l)^2.
\]
Consequently $A=B=0$. The only additional component on the torus therefore comes from a relation involving two entries of the same $v$-row, with no $u$-terms.

It remains to control components on the boundary $u_i=0$. The explicit boundary formula \eqref{eq:boundarydelta} is
\[
 \Delta|_{u_i=0}=-4v_i^2\prod_{j\ne i}u_j
 \left(\sum_{j\ne i}u_j\prod_{k\ne i,j}v_k\right).
\]
The parenthesized cubic is irreducible: it is primitive and linear in its three $u$-variables, and their three coefficient monomials have gcd one. A noncoordinate linear equation $L$, reduced modulo $u_i$, can divide this product only by being proportional to $v_i$ or to one of the $u_j$. Thus boundary factors occur only for a relation in one column or a relation in one row.

Up to interchanging the two rows and permuting columns, this leaves the following alternatives.
\begin{enumerate}
\item For a relation between two entries of the same row, with both coefficients nonzero, the restriction has the form $\ell F_7$, where $F_7$ is irreducible of degree seven and has multiplicity one. For example, put $u_j=\kappa u_i$, $\kappa\ne0$. The coefficient of the lowest power $u_i$ is
\[
 -4v_iv_j(\kappa v_i+v_j)u_ku_l(u_lv_k+u_kv_l),
\]
which is not zero. Hence the boundary factor has multiplicity exactly one; the unique dominating component accounts for the remaining degree seven.
\item For a relation between the two entries of one column, both coefficients nonzero, the restriction has the form $u_i^2F_6$, with $F_6$ irreducible of degree six. Here $z_i$ is fixed to a nonzero constant. The remaining specialization of $q$ is irreducible: homogenizing a putative nontrivial factorization recovers a factorization of the homogeneous irreducible $q$. It has degree four, since its part of degree four is not zero. The boundary formula, or the direct substitution $v_i=\kappa u_i$, gives precisely the square factor $u_i^2$ and no additional boundary factor.
\item In every other noncoordinate case the restriction is irreducible of degree eight. There is one reduced dominating component and no boundary or vertical component.
\end{enumerate}
Every alternative has an irreducible factor of odd multiplicity and degree greater than four. Equation~\eqref{eq:oddobstruction} therefore excludes all noncoordinate hyperplanes.

\subsection{The content of an isotropic frame}\label{sec:spinorcontent}
The remaining hyperplane is the coordinate one. Its determinant has an odd cubic factor, which is too small for the preceding argument. We need the content of the maximal minors of a linear isotropic four-frame in a hyperbolic space of dimension eight.

On an open set where the upper block $A$ is invertible, write the frame as $\binom{I}{S}A$, with $S$ skew-symmetric. Consider the eight rational coordinates
\[
 (1,S_{12},S_{13},S_{14},S_{23},S_{24},S_{34},\Pf S).
\]
Clear homogeneous denominators and divide by the common gcd to obtain a primitive homogeneous polynomial vector $p$ of some common degree $e$. The coordinates have equal degree because their rational ratios are homogeneous of degree zero.

Every $4\times4$ minor of $\binom{I}{S}$ is a quadratic expression in these eight coordinates. With zero lower rows the minor is $1$. With one or two lower rows it is respectively an entry of $S$ or a $2\times2$ minor. With three lower rows the corresponding $3\times3$ minor of a $4\times4$ skew matrix is $\Pf(S)$ times an entry of $S$, with the appropriate sign. With four lower rows it is $\Pf(S)^2$. These identities follow by expanding the determinants, or from the Pfaffian formula in size four. Among the minors are the squares of all eight coordinates: the empty minor, the six principal $2\times2$ minors, and the full determinant.

It follows that the maximal minors of the original frame are $g$ times these quadratic expressions in $p$, where initially $g=\det(A)/p_0^2$ is rational. The quadratic expressions have gcd one because they include every coordinate square of a primitive vector. Taking valuations at each irreducible polynomial shows that $g$ is polynomial and is exactly the gcd of the maximal minors, up to a nonzero constant. In particular,
\begin{equation}\label{eq:evengcd}
 \deg g=4-2e\in\{0,2,4\}.
\end{equation}
For two frames, the Pfaffian expansion also gives
\begin{equation}\label{eq:spinfactor}
 \det(B|_K)=g_Ag_C h^2,
\end{equation}
where $h$ is polynomial. More explicitly, $h=p_0q_0\Pf(T-S)$ is a bilinear polynomial in their eight homogeneous coordinates: expand the Pfaffian of the difference of two $4\times4$ skew matrices.

If an irreducible cubic divides this determinant to odd multiplicity, it divides one of $g_A,g_C$, and~\eqref{eq:evengcd} forces that gcd to have degree four. Then $e=0$, so all Pl\"ucker coordinates of that frame are proportional to constants, and its span is one constant Lagrangian. To see the last assertion, take a nonzero maximal minor, express the remaining rows by Cramer's rule in the four selected rows, and observe that all the resulting ratios are constant. The frame lies in a fixed four dimensional space. The cross pairing consequently factors through four linear channels.

\subsection{The coordinate hyperplane}\label{sec:coordK7}
Put $u_4=0$, write $u,v$ for the first three entries in the two rows, and put $t=v_4$. With $C=C(u)$ as in Section~\ref{sec:targetfacts}, set $w=Cv$ and
\[
 R=u_1v_2v_3+u_2v_1v_3+u_3v_1v_2.
\]
This cubic is irreducible because it is primitive and linear in $u$, with coefficient monomials of gcd one. Directly,
\begin{equation}\label{eq:B7}
 B_7=\begin{pmatrix}tC&w\\w^T&0\end{pmatrix},\qquad
 \det B_7=-t^2\det(C)v^TCv=-4u_1u_2u_3t^2R.
\end{equation}
By the preceding subsection, the odd cubic factor $R$ forces a factorization $AF=B_7$ with $A,F$ linear $4\times4$ matrices.

At $t=0$ the matrix $B_7$ has rank two. Its row space and column space are spanned by $e_4$ and $(w,0)$. The three coordinates of $w$ are primitive quadrics: their gcd is one. If $\det A$ were not divisible by $t$, then $A|_{t=0}$ would be invertible over the rational function field in $u,v$, and every row of $F|_{t=0}$ would have its first three coordinates a rational multiple of $w$. A linear vector cannot be a nonzero rational multiple of a primitive quadratic vector. Indeed, primitivity makes that multiplier polynomial, and degree then makes it zero. All those rows would be multiples of $e_4$, contradicting rank two. Transposing the argument proves that $\det F$ is also divisible by $t$.

The determinant in \eqref{eq:B7} has $t$-valuation two, so each factor has valuation one. The cubic $R$ divides one factor. Interchanging and transposing the factors if necessary, degree four gives
\[
 \det A=\alpha tR,\qquad \det F=\beta u_1u_2u_3t,
\]
with $\alpha,\beta\ne0$.

On the hyperplanes $u_1=0,u_2=0,u_3=0$, $B_7$ has respectively the kernel vectors
\[
 (0,u_2,-u_3,0)^T,\qquad
 (u_1,0,-u_3,0)^T,\qquad
 (u_1,-u_2,0,0)^T.
\]
The matrix $A$ is generically invertible on each of these hyperplanes, so $F$ kills those vectors. For every row $(f_1,f_2,f_3,f_4)$ of $F$,
\[
 u_2f_2-u_3f_3\in(u_1),\quad
 u_1f_1-u_3f_3\in(u_2),\quad
 u_1f_1-u_2f_2\in(u_3).
\]
Comparison of coefficients first kills every $v$- and $t$-coefficient of $f_1,f_2,f_3$. Comparing the remaining quadratic monomials in $u$ gives
\[
 f_1=a u_2+b u_3,\quad
 f_2=a u_1+c u_3,\quad
 f_3=b u_1+c u_2
\]
for constants $a,b,c$. This is a constant row times $C(u)$. Thus the first three columns of $F$ equal $TC(u)$ for a constant $4\times3$ matrix $T$, necessarily of rank three.

The first three columns of $AF=B_7$ now give, after cancellation of the generically invertible $C$,
\[
 AT=\binom{tI_3}{v^T}.
\]
A constant internal basis change makes $T=\binom{I_3}{0}$. The factors have the forms
\[
 F=\begin{pmatrix}C&p\\0&\ell\end{pmatrix},\qquad
 A=\begin{pmatrix}tI_3&a\\v^T&\alpha_0\end{pmatrix}.
\]
Since $\det C=2u_1u_2u_3$, the determinant of $F$ forces $\ell=(\beta/2)t$. But the upper three entries of the last column of $AF$ are $tp+a\ell$, divisible by $t$, whereas the corresponding entries of $B_7$ are $w=Cv$, a nonzero vector independent of $t$. This contradiction excludes the coordinate hyperplane.

This excludes the space of two rows and all its hyperplanes. With Section~\ref{sec:small}, the only admissible kernel left is a full row.

\section{The full row}
\label{sec:fullrow}
The full row is the last case. Here $h=13$ and $a=12$, so $E$ has dimension two, and the Hessian vanishes on $K$. Section~\ref{sec:invariant13} reduces a representation of size $14$ with this kernel to a nilpotent pencil $S(z)=\sum_jz_jS_j$ on a space $V=U\oplus\C e$ of dimension $13$, together with symmetric matrices $C_Y$ linear in $Y$. These are subject to three identities, which together express $F_z(Y)$ through $S$ and $C$. We then show that this system has no solution, without splitting into cases by the type of the pencil. The pencil couples the lower rows to the extra direction $e$ through a covector $\ell$. When $\ell$ involves two or more rows, the kernels of $\ell$ inside the rows are eigenspaces of the whole pencil, so nilpotence forces the eigenvalues to vanish. What remains is $S(Y,0)=\ell(Y)(v,1)$ for a vector $v$, and one of the three identities then forces $\ell$ to divide $F_z$, which is impossible. When the coupling sits in a single row, the rows $b$ and $c$ of $Y$ are invariant under the whole pencil, so its compressions to them are nilpotent. The covectors on which the pencil can then couple form ten lines and fifteen isolated points, and the linearity of the pencil in $z$ leaves four types. A trace computation removes the eigenvalue for each of the four covector types, and at an explicit endpoint where the remaining action vanishes one finds $F_z(Y)=1$, $-2$ or $-4$, a contradiction. For the fourth type $\ell$ divides $F_z$ again.

Throughout this section the endpoint kernel is the first row, written $z=(z_1,z_2,z_3,z_4)$, and $Y=(a,b,c)$ denotes the three lower rows. The endpoint image $U$ has dimension twelve and $W$ has dimension twenty-six.

\subsection{An invariant subspace of dimension thirteen}\label{sec:invariant13}
The Hessian vanishes at every point of the row. Therefore the image of each $R_z$ is isotropic in the nondegenerate two dimensional space $E$, and has dimension at most one. There are exactly two isotropic lines in that space. The conditions that the entire image of a matrix lie in either fixed line are polynomial closed conditions on its entries. The family $z\mapsto R_z$ is polynomial and its parameter space is irreducible. Hence either $R_z=0$ for all $z$, or every image lies in one fixed isotropic line $L\subset E$.

The first case is impossible. It makes $U$ invariant under all $M_z$. For $z$ of full support, Lemma~\ref{lem:cubicresolvent} gives a nilpotent infinitesimal stabilizer of $F_z$ on $U$. By Lemma~\ref{lem:fullstab} it is scalar on each row, hence zero, and Lemma~\ref{lem:cubicresolvent} then gives $F_z=0$, a contradiction.

In the second case, choose once and for all a lift $e\in U^\perp$ of $L$, and put $V=U\oplus\C e$. This is a maximal isotropic subspace of dimension thirteen, and $M_zU\subseteq V$. We claim $M_zV\subseteq V$ for every $z$.

Self-adjointness first gives $M_ze\in U^\perp$, because
\[
 \langle M_ze,u\rangle=\langle e,M_zu\rangle=0
 \qquad(u\in U),
\]
with both $e$ and $M_zu$ in the isotropic space $V$. For a generic $z$ with $R_z\ne0$, choose $u\in U$ for which
\[
 M_zu=u_0+\alpha e,\qquad u_0\in U,\quad\alpha\ne0.
\]
The third moment in \eqref{eq:allmoments} gives
\[
 0=\langle u,M_z^3u\rangle
  =\langle M_zu,M_z^2u\rangle
  =\alpha^2\langle e,M_ze\rangle.
\]
For the last equality, expand $M_z^2u=M_zu_0+\alpha M_ze$. The term pairing $u_0$ with $M_zu_0$ and the term pairing $e$ with $M_zu_0$ vanish in $V$; the term pairing $u_0$ with $M_ze$ vanishes because $M_ze\in U^\perp$. Thus $\langle e,M_ze\rangle=0$. In $E$, the orthogonal complement of an isotropic line is that same line. It follows that $M_ze\in V$. This holds on a dense set of $z$ and extends to all $z$ by polynomiality, so $V$ is invariant under every $M_z$.

Use the coordinates on $U$ given by $\Phi$ on the twelve variables of the lower rows, together with the fixed vector $e$. Let $S(z)$ be $M_z|_V$. It is linear in all four $z_j$, and $S(z)^{13}=0$ for every $z$, since it is the restriction of the nilpotent $M_z$.

For $Y\in U$, define a symmetric matrix $C_Y$ on $V$ by restricting the bilinear form
\[
 (v,w)\longmapsto\langle v,M_Yw\rangle.
\]
It is symmetric by self-adjointness of $M_Y$ and linear in $Y$. With $y=(Y,0)\in V$, the part of the determinant identity of degree three in $Y$ reads
\begin{equation}\label{eq:fullresolvent}
 \frac12\bigl((I-S(z))^{-1}y\bigr)^TC_Y
       \bigl((I-S(z))^{-1}y\bigr)=F_z(Y).
\end{equation}
To verify~\eqref{eq:fullresolvent}, let $R=(I-D_z)^{-1}$. At $Y=0$, $\det(I-D_z)=1$. The part of degree two in $Y$ is $b(Y)Rc(Y)=0$, because every term of $f$ uses exactly three variables of the lower rows. In degree three, the term coming from the determinant factor multiplies this zero quadratic, which leaves $b(Y)RD_YRc(Y)=F_z(Y)$. In $W$ this is one half of the quadratic form in~\eqref{eq:fullresolvent}, and invariance of $V$ identifies the resolvent with $(I-S(z))^{-1}$.

Replacing $z$ by $tz$ and comparing coefficients up to $t^2$ gives the conditions
\begin{align}
 S(z)^{13}&=0, &&\tag{N}\label{eq:N}\\
 y^TC_Yy&=0, &&\tag{H0}\label{eq:H0}\\
 y^TC_YS(z)y&=F_z(Y), &&\tag{H1}\label{eq:H1}\\
 y^TC_YS(z)^2y+\tfrac12(S(z)y)^TC_Y(S(z)y)&=0.
 &&\tag{H2}\label{eq:H2}
\end{align}
We will show that this system alone has no solution; the higher coefficients are not needed.

\subsection{The quotient-stabilizer identity}\label{sec:universalcorrection}
Fix $z$ for the moment and write, in the decomposition $V=U\oplus\C e$,
\[
 S=\begin{pmatrix}A&d\\\ell&\sigma\end{pmatrix},\qquad
 C_Y=\begin{pmatrix}Q_Y&q_Y\\q_Y^T&c_Y\end{pmatrix}.
\]
We call the coordinate along $e$ the hidden coordinate. The compression $A$ of $S$ to $U$ need not be nilpotent, and $d$ and $\sigma$ may be nonzero. We show that~\eqref{eq:H0}--\eqref{eq:H2} imply
\begin{equation}\label{eq:quotientcondition}
 \delta_AF_z\in(\ell).
\end{equation}
Here $(\ell)$ is the ideal generated by the linear form $\ell(Y)$; when $\ell=0$ the condition is $\delta_AF_z=0$.

Write $q_Y=MY$ and $c_Y=g^TY$, and extend $Y\mapsto C_Y$ linearly to $V$ by setting
\[
 C_e=\begin{pmatrix}-(M+M^T)&-g/2\\-g^T/2&0\end{pmatrix}.
\]
For $v=(Y,t)$, direct expansion gives $v^TC_vv=0$: the cubic on $U$ is zero by \eqref{eq:H0}; the $t$ coefficient $2Y^TMY$ cancels $Y^T(M+M^T)Y$; and the $t^2$ coefficient $g^TY$ cancels the two off-diagonal $-g/2$ contributions. Define
\[
 G(v)=-\tfrac12v^TC_{Sv}v.
\]
Polarizing the identity $v^TC_vv=0$ in a direction $w$ gives
\[
 v^TC_wv+2w^TC_vv=0.
\]
At $w=Sv$ it follows that $G(v)=v^TC_vSv$. In particular $G|_U=F_z$ by \eqref{eq:H1}. Differentiating the first expression for $G$ in direction $Sv$ gives
\[
 \delta_SG(v)=-\tfrac12\bigl(2(Sv)^TC_{Sv}v+v^TC_{S^2v}v\bigr).
\]
In the expansion of the same identity at $v+sSv$, the coefficient of $s^2$ is
\[
 (Sv)^TC_v(Sv)+2v^TC_{Sv}(Sv)=0,
\]
and its linear coefficient in direction $S^2v$ is $v^TC_{S^2v}v=-2v^TC_vS^2v$. Therefore
\[
 \delta_SG(v)=v^TC_vS^2v+\tfrac12(Sv)^TC_v(Sv).
\]
At $v=y=(Y,0)$, \eqref{eq:H2} makes this zero. But the directional derivative there also equals
\[
 \delta_AF_z(Y)+\ell(Y)\,\partial_tG(Y,0).
\]
This proves~\eqref{eq:quotientcondition}.

Expanding directly gives the more explicit identity
\begin{equation}\label{eq:blockcorrection}
 \delta_A H_1=H_2+\ell(Y)
 \left(q_{AY}^TY-Y^TQ_Yd-\sigma q_Y^TY
                 -\tfrac12c_Y\ell(Y)\right).
\end{equation}
Here $q_{AY}$ is the linear map $Y\mapsto q_Y$ evaluated at $AY$. This form of the identity is the one checked in archive~B.

\subsection{Solving the quotient-stabilizer equation}\label{sec:quotienttheorem}
For $z$ of full support, the column scaling in the proof of Lemma~\ref{lem:fullstab} reduces the cubic to
\[
 F(a,b,c)=\sum_{i,j,k\text{ distinct}}a_ib_jc_k=b^TB_1(a)c.
\]
Scalar multiples of $F$ do not affect the condition \eqref{eq:quotientcondition}. Write
\[
 \ell=\alpha(a)+\beta(b)+\gamma(c),\qquad \delta_AF=\ell\Psi,
\]
where $\Psi$ is homogeneous quadratic. We show the following.
\begin{equation}\label{eq:multirowform}
 \begin{gathered}
 \text{If at least two of }\alpha,\beta,\gamma\text{ are nonzero,}\qquad
 A=v\ell+\diag(s_aI_4,s_bI_4,s_cI_4),\quad s_a+s_b+s_c=0;
 \end{gathered}
\end{equation}
and, if $\ell$ is supported on the $a$ row,
\begin{equation}\label{eq:onerowform}
 A=\begin{pmatrix}P&0&0\\v_b\ell&Q&0\\v_c\ell&0&R\end{pmatrix}.
\end{equation}
Both describe $A$ itself, in the decomposition of $U$ by rows.

First we remove the parts of $\Psi$ that involve a single row. If $\alpha\ne0$, the $a^3$ term of the equation gives $\alpha\Psi_{aa}=0$, so $\Psi_{aa}=0$. If $\alpha=0$, choose a nonzero other component, say $\beta$. The $a^2b$ coefficient is
\[
 B_1(a)Pa=\beta_{\mathrm{vec}}\Psi_{aa}(a),
 \qquad P=A_{c\leftarrow a}.
\]
Multiply by $\adj B_1(a)$. If $\Psi_{aa}\ne0$, the irreducible quartic $\det B_1$ cannot divide that quadratic. It must divide each cubic entry of $(\adj B_1)\beta_{\mathrm{vec}}$, so every such entry is zero. Generic invertibility of $B_1$ and $\beta\ne0$ contradict this. Thus $\Psi_{aa}=0$ also in this case. Applying the same argument with each row in the repeated position removes $\Psi_{bb}$ and $\Psi_{cc}$.

Now the $a^2b$ equation is
\[
 F(a,b,Pa)=\alpha(a)\Psi_{ab}(a,b),\qquad P=A_{c\leftarrow a}.
\]
If $\alpha=0$, generic invertibility of $B_1(a)$ gives $P=0$. Otherwise restrict to $a\in\ker\alpha$. The irreducible quartic determinant is not divisible by the linear form $\alpha$, so $B_1(a)$ is invertible on a dense subset of that hyperplane. Hence $Pa=0$ there and, by linearity, on the entire hyperplane. Thus $P=v_c\alpha$ for a constant vector $v_c$. Cancelling $\alpha$ gives $\Psi_{ab}=F(a,b,v_c)$. If $\beta\ne0$, the $b^2a$ equation determines the same vector $v_c$, because the map $v_c\mapsto F(a,b,v_c)$ is injective. The other row pairs give all the remaining off-diagonal blocks by the identical coefficient rule with the three row labels permuted.

If at least two row components of $\ell$ are nonzero, each of the three parts of $\Psi$ that mix two rows is determined in this way. There is a vector $v=(v_a,v_b,v_c)$ with $\Psi=\partial_vF$. Subtracting $v\ell$ from $A$ leaves an infinitesimal stabilizer of $F$. Lemma~\ref{lem:fullstab} proves \eqref{eq:multirowform}. If only $\alpha$ is nonzero, the only possible off-diagonal blocks are $A_{b\leftarrow a}=v_b\alpha$ and $A_{c\leftarrow a}=v_c\alpha$, proving \eqref{eq:onerowform}.

\subsection{Invariant rows and nilpotence}\label{sec:invariantrows}
If $W_0\subseteq U$ satisfies $W_0\subseteq\ker\ell$ and $A(W_0)\subseteq W_0$, then $W_0$ is invariant under $S$ itself: for $w\in W_0$,
\[
 S(w,0)=(Aw,0).
\]
Consequently $A|_{W_0}$ is nilpotent. This is the only way in which nilpotence passes to $A$ below.

\begin{remark}
The compression $A$ itself need not be nilpotent. The matrix
\[
 \begin{pmatrix}1&1\\-1&-1\end{pmatrix}
\]
has square zero, while its compression to the first coordinate is $[1]$. Archive~B contains an affine pencil, whose determinant is a reducible quartic, in which this happens. For this reason nilpotence is only ever applied to restrictions to invariant subspaces.
\end{remark}

If $\ell=0$, then $U$ itself is invariant, \eqref{eq:quotientcondition} says that $A$ stabilizes $F$, and Lemma~\ref{lem:fullstab} makes $A$ scalar on each row. Nilpotence kills all three scalars. Then \eqref{eq:H1} gives $F=0$, impossible.

If $\ell$ has at least two nonzero row components, use \eqref{eq:multirowform}. In each row, the kernel of the corresponding covector has dimension at least three and consists of eigenvectors with eigenvalue $s_a$, $s_b$ or $s_c$. Each of these subspaces is invariant under $S$. Nilpotence forces all three scalars zero, giving $A=v\ell$. Then
\[
 S(Y,0)=\ell(Y)(v,1),\qquad
 F(Y)=\ell(Y)\,y^TC_Y(v,1),
\]
a factorization of the irreducible cubic into a linear and a quadratic factor. So at every point $z$ of full support, $\ell$ is nonzero and supported on a single row.

This row does not depend on $z$. Indeed $z\mapsto\ell_z$ is linear before the column normalization, and its values on a dense open subset lie in the union of the three subspaces of covectors supported on one row. The image of an irreducible space lies in one member of this finite union of closed sets. After a fixed permutation of the lower rows, we may assume that $\ell_z$ is supported on the $a$ row for every $z$.

In~\eqref{eq:onerowform}, the $b$ and $c$ rows are invariant subspaces of $S$, so $Q$ and $R$ are nilpotent. Nothing of the kind is claimed for $P$ or for $A$.

\subsection{Exceptional covectors}
On $\ker\ell$ in the $a$ row, the quotient-stabilizer identity becomes
\[
 B_1(Pa)+Q^TB_1(a)+B_1(a)R=0.
\]
Put $\overline R=(Q+R)/2$ and $K_0=(Q-R)/2$. Taking symmetric and skew parts gives
\begin{align}
 B_1(Pa)+\overline R^TB_1(a)+B_1(a)\overline R&=0,
 \label{eq:symmetricexception}\\
 K_0^TB_1(a)&=B_1(a)K_0
 \label{eq:skewexception}
\end{align}
for all $a\in\ker\ell$. If both $\overline R$ and $K_0$ were scalar, then $Q,R$ would be scalar and their nilpotence would make them zero. Injectivity of $a\mapsto B_1(a)$ would give $Pa=0$ on $\ker\ell$, so the whole $A$ would be $v\ell$, already excluded. Hence at least one of the two equations has a nonscalar solution. We now determine the covectors for which this happens.

\subsection{The ten exceptional lines}\label{sec:tenlines}
In this calculation we index coordinates by $0,1,2,3$, and write the covector as $(r_0,r_1,r_2,r_3)$. Equation \eqref{eq:skewexception} is equivalent to
\[
 K_0^TB_1(a)-B_1(a)K_0=\ell(a)Z
\]
for a constant skew matrix $Z$. Put $\zeta_i=\sum_j Z_{ij}$. Comparing the coefficients obtained at $a=e_i$ gives, for $j\ne i$,
\[
 (K_0)_{ji}=r_i\bigl(\zeta_i/2-Z_{ij}\bigr).
\]
Subtract a scalar matrix and set $(K_0)_{00}=0$. For $q\ne0$, let $i,t$ be the two remaining indices, in increasing order. The diagonal entries are then forced to be
\[
 (K_0)_{qq}=-r_iZ_{0q}+(K_0)_{t0}-(K_0)_{tq}.
\]
Substitute these expressions into all twenty-four coefficient equations: four choices of $a=e_i$ and six entries above the diagonal in the skew identity. Their constant row span, considering coefficients in the $r_i$ and the six entries of $Z$, has dimension nine. Selecting the rows with indices $3,4,5,11,14,16,18,19,21$ gives a complete system.

Normalize $r_0=1$ and set $(p,q,r)=(r_1,r_2,r_3)$. In the order $(Z_{01},Z_{02},Z_{03},Z_{12},Z_{13},Z_{23})$, the resulting matrix is
\begin{equation}\label{eq:ninebysix}
 M(p,q,r)=
 \begin{pmatrix}
 -q&p&0&-1&0&0\\
 -q&1&p-1&-p&p-1&0\\
 0&1-p&p-1&q&-r&q+r-1\\
 0&1-p+q&p-r-1&0&0&q+r-p\\
 -1&1&p-q&0&-r&r\\
 p-q&1&p-r-1&0&p-q&r\\
 q-r&-1&1&p&-p&0\\
 -1&p-r&1&-q&0&-q\\
 p-q&p-q&0&-r&p&-q
 \end{pmatrix}.
\end{equation}
A nonscalar $K_0$ exists exactly when this matrix has a nonzero kernel: $Z=0$ gives only the removed scalar, and the formulas above recover $K_0$ from every kernel vector $Z$.

There are $\binom96=84$ maximal minors. The certificates use the minors of $2M$, each of which is $64$ times the corresponding minor of $M$. Order the subsets of six rows lexicographically, number them $0,\ldots,83$, and denote the minors of $2M$ by $D_j$.

The coefficient lists in \file{fullrow_attempt/root_ten_line_certificate.json} in archive~B give
\begin{equation}\label{eq:squarecert}
 g_i^2\in(D_0,\ldots,D_{83})\ (i=1,2,3),\qquad
 h_i^2\in(D_0,\ldots,D_{83})\ (i=1,2),
\end{equation}
where
\begin{align*}
 g_1&=(p-1)(q-r)(p+qr),\\
 g_2&=(q-1)(p-r)(q+pr),\\
 g_3&=(r-1)(p-q)(r+pq),\\
 h_1&=(p-q)(1-r)(1-r+r^2-p^2-pq-q^2),\\
 h_2&=(p-r)(1-q)(1-q+q^2-p^2-pr-r^2).
\end{align*}
These are polynomial identities over $\Q$. Each square is a sum of twenty-three polynomial multiples of the $D_j$, and Appendix~\ref{sec:polynomialappendix} lists every multiplier, so the identities can be checked by expanding them.

So the five polynomials vanish at every complex point where $M$ drops rank. We solve these equations. If some coordinate equals one, say $p=1$, the equations $g_2=g_3=0$ imply $q=1$, or $r=1$, or $q+r=0$. These give three of the listed lines below. Permuting $p,q,r$ covers every case with a coordinate equal to one.

Assume none equals one. If two agree, say $p=q=t$, the first two equations reduce to $t(t-1)(t-r)(1+r)=0$. Thus $t=0$, $r=t$, or $r=-1$. The latter two are listed lines, and $t=0$ is a coordinate axis. Permuting covers all cases with two equal coordinates. If no two coordinates agree, all three remaining factors $p+qr$, $q+pr$, $r+pq$ must vanish. Subtracting the first two gives $(p-q)(1-r)=0$, contrary to the assumptions. There are no further cases.

On each of the three extra axes, one of $h_1,h_2$ restricts to $\pm(t-t^3)$. Its roots $0,1,-1$ already belong to the lines. Thus the full affine solution set is
\[
\begin{gathered}
 (t,t,t),\quad(t,1,1),\quad(1,t,1),\quad(1,1,t),\\
 (-1,t,t),\quad(t,-1,t),\quad(t,t,-1),\\
 (t,-t,1),\quad(t,1,-t),\quad(1,t,-t).
\end{gathered}
\]
Conversely each of the eighty-four minors vanishes identically on each displayed line, which amounts to $840$ polynomial identities in one variable. So $M$ drops rank exactly on these ten lines. By coordinate permutations, the chart $r_0\ne0$ covers every nonzero covector. Homogeneously the locus is the union of four projective star lines
\[
 (\alpha,\beta,\beta,\beta)
\]
and their choices of distinguished coordinate, and six projective difference lines
\[
 (\alpha,-\alpha,\beta,\beta)
\]
with the distinguished opposite pair chosen in all six ways.

\subsection{Fifteen further points}\label{sec:normalizer}
It remains to classify the nonscalar $R_0$ such that
\begin{equation}\label{eq:normalizercondition}
 R_0^TB_1(a)+B_1(a)R_0\in\im B_1
 \qquad\text{for every }a\in\ker\ell.
\end{equation}
This condition follows from~\eqref{eq:symmetricexception}.

Again use indices $0,1,2,3$, normalize $\ell_0=1$, and subtract $(R_0)_{00}I$. The diagonal entry with index $i$ of the matrix in \eqref{eq:normalizercondition} is
\[
 2\sum_{j\ne i}B_1(a)_{ij}(R_0)_{ji}.
\]
It must vanish on $\ker\ell$, so its linear coefficient form is a multiple of $\ell$. It is independent of $a_i$. If $\ell_i\ne0$, that multiple is zero, and the invertible system of pair sums on the other three coordinates makes all off-diagonal entries in column $i$ zero. If $\ell_i=0$, solving that system gives
\begin{equation}\label{eq:normalizercolumns}
 (R_0)_{ji}=t_i(E-2\ell_j)\quad(j\ne i),
 \qquad E=\sum_j\ell_j,
\end{equation}
with one free scalar $t_i$, into which a factor two has been absorbed.

A symmetric matrix $C$ with zero diagonal belongs to $\im B_1$ exactly when
\begin{equation}\label{eq:Blinearrelations}
 C_{01}+C_{23}-C_{02}-C_{13}=0,\qquad
 C_{01}+C_{23}-C_{03}-C_{12}=0.
\end{equation}
The two equations follow directly from the entries of $B_1$, which are sums over complementary pairs, and they are independent; their solution space has dimension four, equal to the dimension of the injective image of $B_1$. Evaluate these two equations on the three vectors $a=e_j-\ell_je_0$, $j=1,2,3$. Together with \eqref{eq:normalizercolumns}, these six linear equations make up the whole system. We solve it according to the number of zero coordinates of $\ell$.

\paragraph{No zero coordinate.}
All off-diagonal entries of $R_0$ vanish. Write $R_0=\diag(0,d_1,d_2,d_3)$. The two forms in \eqref{eq:Blinearrelations} are
\[
 L_1=(d_2-d_1)(a_0-a_3)+d_3(a_1-a_2),
 \qquad
 L_2=(d_3-d_1)(a_0-a_2)+d_2(a_1-a_3).
\]
Each must be proportional to $\ell$. If both are nonzero, $\ell$ belongs to the intersection of their coefficient two-planes
\[
 \Span((1,0,0,-1),(0,1,-1,0))
 \cap\Span((1,0,-1,0),(0,1,0,-1)),
\]
which is the line spanned by $(1,1,-1,-1)$. If $L_1=0$ but $L_2\ne0$, then $d_2=d_1,d_3=0$ and $\ell$ is proportional to $(1,-1,-1,1)$. If $L_2=0$ but $L_1\ne0$, it is proportional to $(1,-1,1,-1)$. If both vanish, all $d_i$ vanish and the normalizer is scalar before subtraction. So there are three projective possibilities in this case.

\paragraph{Exactly one zero coordinate.}
Put $\ell=(1,p,q,0)$ with $pq\ne0$. The unknowns are $d_1,d_2,d_3$ and the parameter $t$ of the final column in \eqref{eq:normalizercolumns}. If $t=0$, the six equations force every $d_i=0$. For a nonscalar solution scale $t=1$. Three of the six equations solve to
\[
 d_1=2(p-1),\qquad d_2=2(q-1),\qquad
 d_3=1-p-q+4pq-4p^2.
\]
One remaining equation is $4(p^2-q^2)=0$. If $q=p$, another is $4(p^2-1)=0$. If $q=-p$, the other two equations become
\[
 4(p^2-1)(2p+1)=0,\qquad -4(p^2-1)(2p-1)=0.
\]
Adding them gives $8(p^2-1)=0$. Thus in every case $p^2=q^2=1$. Substitution in the displayed solution verifies all four possibilities. Permuting the zero coordinate gives sixteen projective points; four, with the three nonzero entries equal, already lie on the ten lines. The other twelve are the permutations of $(1,1,-1,0)$, up to overall scaling.

\paragraph{Exactly two zero coordinates.}
Put $\ell=(1,p,0,0)$ with $p\ne0$, and let $s,t$ be the two off-diagonal column parameters. Four of the six equations give
\[
 \begin{gathered}
 d_1=2(p-1)(s-t),\qquad d_2=(1-p)s-2t,\\
 d_3=-2s+(1-p)t,\qquad4(p-1)(s-t)=0.
 \end{gathered}
\]
If $p\ne1$, then $s=t$, and another equation is $4(p^2-1)s=0$. A nonscalar solution therefore requires $p=-1$. If $p=1$, a remaining equation gives $s=t$, which is allowed. Hence only the pair sum and pair difference points occur, and both lie on the ten lines.

\paragraph{Exactly three zero coordinates.}
The covector is a coordinate point, already on a star line. No additional point is needed.

Every equation used in these eliminations comes from~\eqref{eq:normalizercolumns} and~\eqref{eq:Blinearrelations}, so the case split is exhaustive. The exceptional locus is therefore contained in the ten projective lines together with fifteen further points: the twelve points of support three with mixed signs, and the three balanced points of full support.

\subsection{The common family}\label{sec:commonfamily}
Return to the family before the column normalization. Its coupling, supported on one row, has the form
\[
 \ell_z(a)=\sum_{j=1}^4 l_j(z)a_j,
\]
where every $l_j$ is homogeneous linear in $z$. After the column normalization $a_j=z_j\widetilde a_j$, its covector coordinates are
\begin{equation}\label{eq:rquadratic}
 r_j(z)=z_jl_j(z).
\end{equation}
They are homogeneous quadratics, and each is divisible by its own $z_j$. This is where the linearity of the pencil in $z$ is used.

On the dense open set where $z$ has full support, a surviving family has nonzero $r(z)$ and its projective image lies in the exceptional locus just classified. The closure of that rational image is irreducible. It is therefore contained in one of the ten projective lines or one of the fifteen additional points.

It cannot be a fixed point having at least three nonzero coordinates. If $r_j=c_jH$ for the nonzero fixed coordinates $c_j$, then $H$ is a polynomial of degree two, since it equals $r_j/c_j$. Equation \eqref{eq:rquadratic} makes it divisible by the product of at least three different $z_j$, impossible unless $H=0$. The latter would make the whole covector zero. This eliminates the fifteen further points, and also any fixed point of support three or four on the lines.

On a star line, three of the $r_j$ agree. Their common quadratic is divisible by three distinct variables, so it is zero. The direction is therefore a coordinate covector.

On a difference line, after permuting columns,
\[
 r_1=-r_2,\qquad r_3=r_4.
\]
Divisibility and degree give constants $c,d$ such that
\begin{equation}\label{eq:familydifference}
 (r_1,r_2,r_3,r_4)
 =(cz_1z_2,-cz_1z_2,dz_3z_4,dz_3z_4).
\end{equation}
If $d=0$ this is a pair difference; if $c=0$ it is a pair sum. If $cd\ne0$, the ratio $dz_3z_4/(cz_1z_2)$ is nonconstant on the torus where every $z_j\ne0$, so the projective image is dense in that difference line. In this case it suffices to compute the quotient kernel at a generic point of the line, because the family reaches a dense open subset of it.

Four situations remain: a coordinate covector, a pair difference, a pair sum, and a generic point of a difference line. At a fixed $z$, scaling the covector or the vector $e$ changes neither the quotient equation nor the argument with invariant rows, so we may use the normalized representatives below. The normalization is made one point $z$ at a time, after the divisibility argument.

\subsection{Quotient kernels for the four remaining types}\label{sec:fourkernels}
We return to indices $1,2,3,4$. Put
\[
 u=(1,-1,0,0)^T,\qquad r=(0,0,1,1)^T,\qquad
 P_0=uu^T,\qquad P_1=ur^T,
\]
and let
\[
 T=\diag(\alpha I_4,\beta I_4,\gamma I_4),
 \qquad\alpha+\beta+\gamma=0.
\]
In the formulas below $p\in\C^{12}$ is arbitrary, so the term $p\ell$ runs over a twelve dimensional space of matrices.

For $\ell=a_3+a_4$, the solution space of $\delta_AF\in(\ell)$ is
\begin{equation}\label{eq:kernelpairsum}
 A=T+p\ell+s\diag(-P_0,P_0,P_0)
       +t\diag(0,P_1,-P_1).
\end{equation}
For $\ell=a_1-a_2$, the solution space is
\begin{equation}\label{eq:kernelpairdiff}
 A=T+p\ell+s\diag(0,-P_0,P_0)
       +t\diag(P_1,-P_1,-P_1).
\end{equation}
For $\ell=a_1$, put
\[
 w=(1,-1,-1,-1)^T,\quad P=we_1^T,
\]
\[
 Q_0=\begin{pmatrix}
 -1&-1&-1&-1\\0&0&1&1\\0&1&0&1\\0&1&1&0
 \end{pmatrix},\qquad
 R_0=\begin{pmatrix}
 0&1&1&1\\0&1&-1&-1\\0&-1&1&-1\\0&-1&-1&1
 \end{pmatrix}.
\]
Then the solution space is
\begin{equation}\label{eq:kernelcoordinate}
 A=T+p\ell+s\diag(0,P,-P)+t\diag(R_0,Q_0,Q_0).
\end{equation}
At a generic point $\ell=\lambda(a_1-a_2)+\eta(a_3+a_4)$ of a difference line, put
\[
 K_0=u(-\lambda u+\eta r)^T.
\]
At a generic point the solution space is
\begin{equation}\label{eq:kernelgeneric}
 A=T+p\ell+\nu\diag(0,K_0,-K_0).
\end{equation}
Substitution shows that these families solve the equation. That they are the whole solution spaces is a rank computation, which we now describe.

Let $Y_0,\ldots,Y_{11}$ enumerate $a,b,c$ in row order. For a nonzero coefficient of $\ell$, solve for that variable and substitute in the cubic $\delta_AF$. The coefficient matrix has one column for every matrix entry $A_{ij}$ and rows indexed by cubic monomials in the eleven remaining variables. Its column $(i,j)$ is the coefficient vector of
\begin{equation}\label{eq:quotientmatrixrule}
 \left.Y_j\frac{\partial F}{\partial Y_i}\right|_{\ell=0}.
\end{equation}
There are $144$ columns, one for each entry of $A$, and no condition is imposed on $A$.

In each of the first three cases the sixteen displayed matrices are independent and lie in the kernel, and a $128\times128$ minor of the coefficient matrix is nonzero. The minors are as follows.
\begin{center}
\begin{tabular}{lrr}\toprule
Covector & Kernel dimension & Selected minor\\\midrule
$a_1$ &16&$-140737488355328$\\
$a_1-a_2$ &16&$18446744073709551616$\\
$a_3+a_4$ &16&$-9007199254740992$\\
$a_1-a_2+2(a_3+a_4)$ &15&$146884487083177444726800384$\\\bottomrule
\end{tabular}
\end{center}
The last minor is $129\times129$, so the generic rank on the difference line is at least $129$. The fifteen matrices in~\eqref{eq:kernelgeneric} satisfy the equation identically in $(\lambda,\eta)$ and are generically independent, so the generic rank is at most $129$. Thus the generic kernel has dimension fifteen. Section~\ref{sec:commonfamily} explains why a family as in~\eqref{eq:familydifference} with $cd\neq0$ reaches this generic set.

Appendix~\ref{sec:minorappendix} lists the rows and columns of the four minors. The same indices, together with the kernel bases, are stored in \file{fullrow_attempt/system_exceptional_killer_certificate.json} in archive~B, whose checker rebuilds the coefficient matrix, verifies each inclusion and the independence, and evaluates the minors.

\subsection{The terminal contradictions}\label{sec:terminal}
If $S(Y,0)=0$, then~\eqref{eq:H1} forces $F(Y)=0$. We violate this implication in each of the three special types, and find a forbidden factorization in the generic type. The column $d$ and the corner $\sigma$ play no role, because the inputs we test have hidden coordinate zero.

\paragraph{Pair sum.}
In~\eqref{eq:kernelpairsum}, the $b$ and $c$ blocks are
\[
 Q=\beta I+sP_0+tP_1,\qquad R=\gamma I+sP_0-tP_1.
\]
Both nonscalar parts have image in $\C u$. Since $u^Tu=2$ and $r^Tu=0$, their eigenvalues are respectively
\[
 \beta,\beta,\beta,\beta+2s,
 \qquad
 \gamma,\gamma,\gamma,\gamma+2s.
\]
These are restrictions of $S$ to invariant rows, so they are nilpotent. Thus $\beta=\gamma=s=0$, and $\alpha=0$. The remaining operator is $A=p\ell+tN$ with $N=\diag(0,P_1,-P_1)$.

Take
\[
 a=e_1,\qquad b=c=e_3-e_4.
\]
The covector $\ell=a_3+a_4$ vanishes, and $r^Tb=r^Tc=0$, so $NY=0$. Hence $AY=0$ and $S(Y,0)=0$. But $F(Y)=-2$: with $a_1=1$, the only nonzero matching terms use $(b_3,c_4)$ or $(b_4,c_3)$, and each equals $-1$. This contradicts \eqref{eq:H1}.

\paragraph{Pair difference.}
In \eqref{eq:kernelpairdiff}, the invariant row eigenvalues are
\[
 \beta,\beta,\beta,\beta-2s,
 \qquad
 \gamma,\gamma,\gamma,\gamma+2s.
\]
Nilpotence again makes $T=s=0$. Take
\[
 a=e_1+e_2,\qquad b=c=e_3-e_4.
\]
Now $\ell=a_1-a_2=0$. Also $r^Ta=r^Tb=r^Tc=0$, so the remaining block $\diag(P_1,-P_1,-P_1)$ kills $Y$. Hence $S(Y,0)=0$. For each of $a_1,a_2$ there are two nonzero terms of value $-1$, giving $F(Y)=-4$. This is again impossible.

\paragraph{Coordinate covector.}
In \eqref{eq:kernelcoordinate}, $Q_0w=2w$ and $Q_0$ acts as $-I$ on the quotient $\C^4/\C w$. The matrix $P$ has rank one and image $\C w$, and $Pw=w$. The invariant $b$ block therefore has eigenvalues
\[
 \beta-t,\beta-t,\beta-t,\beta+s+2t,
\]
and the invariant $c$ block has
\[
 \gamma-t,\gamma-t,\gamma-t,\gamma-s+2t.
\]
Nilpotence gives $\beta=\gamma=t$, $s+3t=0$, and $-s+3t=0$. Over $\C$ these imply $s=t=0$ and hence $T=0$. Thus $A=p\ell$. Take $a=e_2,b=e_3,c=e_4$. The covector $a_1$ vanishes, so $S(Y,0)=0$, while the single nonzero matching term gives $F(Y)=1$. This is the third contradiction.

\paragraph{Generic difference line.}
In \eqref{eq:kernelgeneric}, the matrix $K_0$ has rank one and trace $-2\lambda$. The invariant $b$ block has eigenvalues $\beta$ three times and $\beta-2\lambda\nu$ once; the $c$ block has $\gamma$ three times and $\gamma+2\lambda\nu$ once. At a generic point with $\lambda\ne0$, nilpotence gives $\beta=\gamma=\nu=0$, and then $\alpha=0$. Again $A=p\ell$. Equation \eqref{eq:H1} factors $F$ by $\ell$, contradicting irreducibility. A family as in~\eqref{eq:familydifference} with $cd\neq0$ reaches such a generic point.

This excludes every case left by the preceding subsections, and proves the following.
\begin{theorem}\label{thm:fullrowsystem}
There are no data $S(z)\in\End(\C^{13})$ linear in $z\in\C^4$ and $C_Y\in\Sym^2((\C^{13})^*)$ linear in the twelve variables of the lower rows, with $y=(Y,0)$, satisfying \eqref{eq:N}--\eqref{eq:H2} identically.
\end{theorem}
By Section~\ref{sec:invariant13}, a representation of size $14$ whose endpoint kernel is a full row would supply such data. So a full row is not the endpoint kernel of any representation of size $14$.

\subsection{Unit identities}\label{sec:unitidentities}
The eigenvalue arguments above can also be written as polynomial combinations equal to one, so that each terminal contradiction becomes a single identity. Let $e_2(Q)$ be the second coefficient of the characteristic polynomial, the sum of the principal $2\times2$ minors. It vanishes whenever $Q$ is nilpotent. The letters $L,M$ below stand for the linear combinations of entries of $C_Y$ that appear in~\eqref{eq:H1} at the witness; the identities hold with $L$ and $M$ treated as independent indeterminates.

For the pair sum, the trace equations give $s=-2\beta$, $\gamma=\beta$, $\alpha=-2\beta$. Then $e_2(Q)=-6\beta^2$, and the witness evaluation gives an equation $E=2+\beta L=0$. Direct expansion yields
\[
 1=\frac{2-\beta L}{4}E-\frac{L^2}{24}e_2(Q).
\]
Thus the equations imply $1=0$.

For the pair difference, the trace equations give $s=2\beta$, $\gamma=-\beta$, $\alpha=0$. With $e_2(Q)=-6\beta^2$ and the witness equation $E=4+\beta L=0$,
\[
 1=\frac{4-\beta L}{16}E-\frac{L^2}{96}e_2(Q).
\]

For the coordinate case, put $u=\beta-t$, $v=\gamma-t$. Solving the two trace equations gives
\[
 t=-\frac{2(u+v)}3,\quad s=2(v-u),\quad
 \beta=\frac{u-2v}3,\quad\gamma=\frac{v-2u}3,\quad
 \alpha=\frac{u+v}3.
\]
Here $e_2(Q)=-6u^2$, $e_2(R)=-6v^2$, and the witness equation is $E=1-uL-vM=0$. Put $w=uL+vM$. Then
\[
 \begin{split}
 1={}&(1+w+w^2)E
 -\frac{uL^3+3vL^2M}{6}e_2(Q)\\
 &-\frac{3uLM^2+vM^3}{6}e_2(R).
 \end{split}
\]
Indeed $(1+w+w^2)(1-w)=1-w^3$, and the other two terms add $w^3$.

For the generic difference line, write the $b$-row scalar as $\mu$ and the $c$-row scalar as $-\mu$, after summing the trace equations. Put
\[
 T_Q=4\mu-2\lambda\nu,\qquad e_2(Q)=6\mu^2-6\lambda\mu\nu,
 \qquad E=4+\mu L.
\]
The identity is
\[
 1=\frac{4-\mu L}{16}E-\frac{L^2}{96}e_2(Q)
       +\frac{\mu L^2}{32}T_Q.
\]
It uses no division by $\lambda$. All four identities, together with the trace substitutions and the witness evaluations, are checked by \file{typefree_attempt/zero_unit_checks.py} in archive~B.

\subsection{End of the proof}
We have excluded every kernel in Theorem~\ref{thm:kernellist}, so no representation of size $14$ exists. Padding turns a smaller representation into one of size $14$, so $\dc(f)\geq15$, and Section~\ref{sec:upper} gives $\dc(f)\leq15$. This proves Theorem~\ref{thm:main}.

The proof is written for $n=4$ over $\C$. It uses the one dimensional complement $\C e$, the irreducible quartic $\det B_1$ and the exact exceptional locus, and it says nothing about larger permanents or about border determinantal complexity. In characteristic two the factor $-4$ in Lemma~\ref{lem:crossdet} vanishes, several divisions by two fail, and so do the witness values $-2$ and $-4$, as they must, since the permanent is then a determinant. The coordinate case of Section~\ref{sec:unitidentities} also divides by three. The theorem is stated over $\C$, and the argument has not been checked uniformly in odd characteristic.

\section{Certificates}\label{sec:audit}
Every computational step is exact, and each comes with a checker that runs on its own. The checkers are collected in two archives, both available at \url{https://zenodo.org/records/24124959}.
\begin{center}\small
\begin{tabular}{ll}\toprule
Archive & SHA-256\\\midrule
A: \file{permanent4_exact_certificates.zip} & \texttt{f76999ef9f286b6d88795f5b93a1fe31}\\
 & \texttt{73d0811d0c4720463a78ae085c51e21f}\\
B: \file{per4_type_independent_exclusion.zip} & \texttt{f6c4e772d3a9d3e392b17507510e6768}\\
 & \texttt{9364bdd3302b411e360cc0221f8dd434}\\\bottomrule
\end{tabular}
\end{center}
Each digest is printed on two lines. Archive~A has $67$ files and archive~B has $126$. Archive~B also contains the material of archive~A, under \file{inherited/}. Each archive carries a JSON manifest with the SHA-256 digest of every other file in it, and archive~B has a script, \file{VERIFY_MANIFEST.py}, that checks its manifest.

The programs check the finite computations quoted in Sections~\ref{sec:foundations} to~\ref{sec:fullrow}: the enumeration of coordinate kernels, ranks of explicit integer and rational matrices, and polynomial identities. The reductions that connect these computations to Theorem~\ref{thm:main} are the subject of the text. All programs use exact integer and rational arithmetic and need Python~3 with SymPy. Archive~A is run by \file{RUN_ALL_CHECKS.py}, which calls the fifteen programs in its directory \file{k6_attempt/}, and archive~B by \file{typefree_attempt/RUN_CHECKS.py}, which calls thirteen programs. Some of the checks rewrite their JSON records, so each archive is best extracted into a fresh directory, with its layout intact, and run there. Tables~\ref{tab:programsA} and~\ref{tab:programsB} list the programs.

\begin{longtable}{p{.55\textwidth}p{.38\textwidth}}
\caption{The programs of archive~A, in the directory \file{k6_attempt/}.}\label{tab:programsA}\\
\toprule Program & What it checks\\\midrule\endfirsthead
\toprule Program & What it checks\\\midrule\endhead
\file{equality_checks.py}&Operator equations for the cross kernels.\\
\file{five_channel_certificate.py}&Flattenings and the directions of rank two.\\
\file{k7_coordinate_certificate.py}&The coordinate hyperplane of two rows.\\
\file{k7_gram_checks.py}&Gram determinants and identities for isotropic frames.\\
\file{k8_clifford_checks.py}&Clifford trace and rank, with a positive control.\\
\file{coordinate_kernels_checks.py}&Enumeration of supports and Hessian ranks.\\
\file{rho_zero_checks.py}&Supports of kernels with zero Hessian.\\
\file{other_coordinate_exclusions_checks.py}&Rank obstructions for the smaller kernels.\\
\file{two_by_three_syzygy_checks.py}&Linear and quadratic syzygy spaces.\\
\file{row_kernel_checks.py}&Stabilizers of the contracted cubics.\\
\file{grassmann_chart_checks.py}&Local equations in the charts.\\
\file{chart_jacobian_checks.py}&The four Jacobian minors of Table~\ref{tab:charts}.\\
\file{hook_K6_linear_sections_checks.py}&Controls with linear sections of the complete cross.\\
\file{kernel_controls_checks.py}&Controls for the kernels and the construction.\\
\file{pencil_equations_checks.py}&The representation of size fifteen.\\\bottomrule
\end{longtable}

\begin{longtable}{p{.55\textwidth}p{.38\textwidth}}
\caption{The programs of archive~B. Paths are relative to the archive.}\label{tab:programsB}\\
\toprule Program & What it checks\\\midrule\endfirsthead
\toprule Program & What it checks\\\midrule\endhead
\file{typefree_attempt/root_target_checks.py}&The cubic $F$ and the independence of its derivatives.\\
\file{typefree_attempt/general_correction_checks.py}&The quotient-stabilizer identity.\\
\file{typefree_attempt/general_nilpotent_control_checks.py}&A pencil whose compression is not nilpotent.\\
\file{typefree_attempt/zero_endpoint_checks.py}&Witness evaluations and coefficient identities.\\
\file{typefree_attempt/zero_nilpotence_checks.py}&Characteristic polynomials on invariant rows.\\
\file{typefree_attempt/zero_unit_checks.py}&The four unit identities.\\
\file{typefree_attempt/nonzero_independent_audit_checks.py}&Invariant blocks and the three witnesses.\\
\file{fullrow_attempt/structure_multirow_quotient_checks.py}&The stabilizer system for row scalings.\\
\file{fullrow_attempt/controls_ten_line_audit.py}&The five square identities and the ten lines.\\
\file{fullrow_attempt/structure_symmetric_normalizer_checks.py}&The fifteen further points.\\
\file{fullrow_attempt/system_verify_exceptional_certificate.py}&The four quotient kernels.\\
\file{fullrow_attempt/controls_checks.py}&Controls for the cubic, and representations of larger size.\\
\file{fullrow_attempt/controls_hook_cross_checks.py}&A control case for the cross.\\\bottomrule
\end{longtable}

\section*{Acknowledgements}
The mathematics were assisted by ChatGPT Pro and later refined and simplified with Claude Opus 5.5. The author credits Fable 5.1 for helping crack a missing step along the way.

\appendix

\section{The five polynomial identities}\label{sec:polynomialappendix}
The symbols $D_0,\ldots,D_{83}$ denote the $6\times6$ minors of $2M(p,q,r)$, indexed by the subsets of six rows of $\{0,\ldots,8\}$ in increasing lexicographic order, with the six columns in their displayed order. This fixes each minor, including its sign. The five polynomials $g_1,g_2,g_3,h_1,h_2$ are those in Section~\ref{sec:tenlines}.
For each $f$ we list a positive integer $N_f$ and every nonzero multiplier $H_{f,j}$, which has integer coefficients, in the identity
\[N_f f^2=\sum_{j=0}^{83}H_{f,j}D_j.\]
Multipliers that are not listed are zero, and each identity has twenty-three nonzero multipliers. Expanding the products checks the identities in $\mathbb Z[p,q,r]$.
\subsection{The identity for $g_1$}
Here $N_{g_1}=512$. The nonzero multipliers are as follows.
\begin{align*}
H_{g_1,0}={}&2pr+6p+2q-12r-6
\end{align*}
\begin{align*}
H_{g_1,1}={}&-5pq+2pr-24p-42q-48r^{2}+113r+158
\end{align*}
\begin{align*}
H_{g_1,2}={}&pq+22p+32qr+62q-56r^{2}-109r-90
\end{align*}
\begin{align*}
H_{g_1,3}={}&-4pq-2pr-30p-34q+136r+182
\end{align*}
\begin{align*}
H_{g_1,4}={}&-4pq-24p-34q+132r+166
\end{align*}
\begin{align*}
H_{g_1,6}={}&-pq-6p-6q+r+30
\end{align*}
\begin{align*}
H_{g_1,7}={}&-4pr-8p+8qr-2q-20r^{2}+44r+62
\end{align*}
\begin{align*}
H_{g_1,8}={}&-3pq-18p-28q+139r+136
\end{align*}
\begin{align*}
H_{g_1,9}={}&4pq+24p+32q-132r-176
\end{align*}
\begin{align*}
H_{g_1,10}={}&-2pq-12p-16q-22r-12
\end{align*}
\begin{align*}
H_{g_1,11}={}&-2pr-2p+8qr-2q-56r^{2}-4r+62
\end{align*}
\begin{align*}
H_{g_1,12}={}&2pr+6p+2q-12r-6
\end{align*}
\begin{align*}
H_{g_1,16}={}&2pq+12p+16q-14r-44
\end{align*}
\begin{align*}
H_{g_1,17}={}&-pq-6p-8q+9r+20
\end{align*}
\begin{align*}
H_{g_1,20}={}&-4pq-24p-36q+92r+100
\end{align*}
\begin{align*}
H_{g_1,23}={}&2pq-4pr+4p+8qr+14q+28r^{2}-46r\\
&-98
\end{align*}
\begin{align*}
H_{g_1,24}={}&-4pq-24p-36q+52r+44
\end{align*}
\begin{align*}
H_{g_1,36}={}&24r+56
\end{align*}
\begin{align*}
H_{g_1,38}={}&-16p-32qr-56q+56r^{2}+104r+64
\end{align*}
\begin{align*}
H_{g_1,40}={}&-24r+8
\end{align*}
\begin{align*}
H_{g_1,45}={}&-4p-8qr+56r^{2}+48r+4
\end{align*}
\begin{align*}
H_{g_1,47}={}&8r
\end{align*}
\begin{align*}
H_{g_1,65}={}&-48r-56
\end{align*}
\subsection{The identity for $g_2$}
Here $N_{g_2}=512$. The nonzero multipliers are as follows.
\begin{align*}
H_{g_2,0}={}&-pr+6p+2q-5r^{2}-3r-7
\end{align*}
\begin{align*}
H_{g_2,1}={}&-5pq-26pr+26p+30qr-102q-12r^{3}+54r^{2}\\
&-34r+115
\end{align*}
\begin{align*}
H_{g_2,2}={}&pq+16pr^{2}+5pr-4p+26qr+106q-8r^{3}\\
&-31r^{2}-33r-68
\end{align*}
\begin{align*}
H_{g_2,3}={}&-4pq-19pr+10p+24qr-82q+37r^{2}-5r\\
&+175
\end{align*}
\begin{align*}
H_{g_2,4}={}&-4pq-20pr+16p+24qr-82q+32r^{2}+158
\end{align*}
\begin{align*}
H_{g_2,6}={}&-pq-5pr+4p+6qr-18q+5r^{2}-7r\\
&+24
\end{align*}
\begin{align*}
H_{g_2,7}={}&4pr^{2}+2pr-12p+8qr-2q-8r^{3}+22r^{2}\\
&-26r+40
\end{align*}
\begin{align*}
H_{g_2,8}={}&-3pq-15pr+12p+18qr-64q+31r^{2}+11r\\
&+134
\end{align*}
\begin{align*}
H_{g_2,9}={}&4pq+20pr-16p-24qr+80q-36r^{2}+4r\\
&-168
\end{align*}
\begin{align*}
H_{g_2,10}={}&-2pq-10pr+8p+12qr-40q+4r^{2}+8r\\
&-28
\end{align*}
\begin{align*}
H_{g_2,11}={}&4pr^{2}+pr-6p+8qr-2q-8r^{3}+7r^{2}\\
&-63r+63
\end{align*}
\begin{align*}
H_{g_2,12}={}&-pr+6p+2q-5r^{2}-3r-7
\end{align*}
\begin{align*}
H_{g_2,16}={}&2pq+10pr-8p-12qr+40q-10r^{2}-2r\\
&-28
\end{align*}
\begin{align*}
H_{g_2,17}={}&-pq-5pr+4p+6qr-20q+5r^{2}+r\\
&+14
\end{align*}
\begin{align*}
H_{g_2,20}={}&-4pq-20pr+16p+24qr-84q+20r^{2}+8r\\
&+92
\end{align*}
\begin{align*}
H_{g_2,23}={}&2pq+4pr^{2}+12pr-20p-4qr+38q+4r^{3}\\
&-24r^{2}+28r-88
\end{align*}
\begin{align*}
H_{g_2,24}={}&-4pq-20pr+16p+24qr-84q+12r^{2}+12r\\
&+36
\end{align*}
\begin{align*}
H_{g_2,36}={}&26r+86
\end{align*}
\begin{align*}
H_{g_2,38}={}&-16pr^{2}-32qr-88q+8r^{3}+24r^{2}+60r+60
\end{align*}
\begin{align*}
H_{g_2,40}={}&-8r^{2}+30r+30
\end{align*}
\begin{align*}
H_{g_2,45}={}&-4pr^{2}-8qr+8r^{3}+4r^{2}+60r
\end{align*}
\begin{align*}
H_{g_2,47}={}&6r+2
\end{align*}
\begin{align*}
H_{g_2,65}={}&-12r^{2}-56
\end{align*}
\subsection{The identity for $g_3$}
Here $N_{g_3}=512$. The nonzero multipliers are as follows.
\begin{align*}
H_{g_3,0}={}&-7pr+36p+4q+19r^{2}-7r-37
\end{align*}
\begin{align*}
H_{g_3,1}={}&-10pq-2pr+4p+10q^{2}+80qr-214q+12r^{3}\\
&-76r^{2}-97r+557
\end{align*}
\begin{align*}
H_{g_3,2}={}&2pq+16pr^{2}-33pr+12p-2q^{2}-32qr^{2}+48qr\\
&+246q+24r^{3}-53r^{2}-182r-294
\end{align*}
\begin{align*}
H_{g_3,3}={}&-8pq+11pr-68p+8q^{2}+64qr-172q-43r^{2}\\
&+27r+685
\end{align*}
\begin{align*}
H_{g_3,4}={}&-8pq+4pr-28p+8q^{2}+64qr-172q-32r^{2}\\
&+36r+632
\end{align*}
\begin{align*}
H_{g_3,6}={}&-2pq+pr-12p+2q^{2}+16qr-38q+7r^{2}\\
&-22r+74
\end{align*}
\begin{align*}
H_{g_3,7}={}&4pr^{2}+6pr-72p-8qr^{2}+16qr+4q-6r^{2}\\
&-46r+226
\end{align*}
\begin{align*}
H_{g_3,8}={}&-6pq+3pr-16p+6q^{2}+48qr-134q-51r^{2}\\
&+70r+558
\end{align*}
\begin{align*}
H_{g_3,9}={}&8pq-4pr+32p-8q^{2}-64qr+168q+36r^{2}\\
&-32r-648
\end{align*}
\begin{align*}
H_{g_3,10}={}&-4pq+2pr-16p+4q^{2}+32qr-84q-8r^{2}\\
&-14r-96
\end{align*}
\begin{align*}
H_{g_3,11}={}&4pr^{2}-pr-32p-8qr^{2}+16qr+4q+24r^{3}\\
&-49r^{2}-219r+253
\end{align*}
\begin{align*}
H_{g_3,12}={}&-7pr+36p+4q+19r^{2}-7r-37
\end{align*}
\begin{align*}
H_{g_3,16}={}&4pq-2pr+16p-4q^{2}-32qr+84q+2r^{2}\\
&+16r-120
\end{align*}
\begin{align*}
H_{g_3,17}={}&-2pq+pr-8p+2q^{2}+16qr-42q-r^{2}\\
&-6r+58
\end{align*}
\begin{align*}
H_{g_3,20}={}&-8pq+4pr-24p+8q^{2}+64qr-176q-28r^{2}\\
&+36r+400
\end{align*}
\begin{align*}
H_{g_3,23}={}&4pq+4pr^{2}+4pr-56p-4q^{2}-8qr^{2}-16qr\\
&+88q-12r^{3}+48r^{2}+86r-342
\end{align*}
\begin{align*}
H_{g_3,24}={}&-8pq+4pr-24p+8q^{2}+64qr-176q-28r^{2}\\
&+32r+184
\end{align*}
\begin{align*}
H_{g_3,36}={}&-24r^{2}+50r+198
\end{align*}
\begin{align*}
H_{g_3,38}={}&-16pr^{2}+32pr+32qr^{2}-64qr-208q-24r^{3}+32r^{2}\\
&+228r+220
\end{align*}
\begin{align*}
H_{g_3,40}={}&-8r^{2}+22r-10
\end{align*}
\begin{align*}
H_{g_3,45}={}&-4pr^{2}+8pr-4p+8qr^{2}-16qr-8q-24r^{3}\\
&+44r^{2}+212r+4
\end{align*}
\begin{align*}
H_{g_3,47}={}&-8r^{2}+14r+10
\end{align*}
\begin{align*}
H_{g_3,65}={}&12r^{2}-16r-216
\end{align*}
\subsection{The identity for $h_1$}
Here $N_{h_1}=512$. The nonzero multipliers are as follows.
\begin{align*}
H_{h_1,0}={}&12p^{2}r-12p^{2}+6pr-41p+2qr-7q+12r^{3}\\
&-3r^{2}-31r+52
\end{align*}
\begin{align*}
H_{h_1,1}={}&-8p^{2}r+8p^{2}+15pq-24pr+94p-10q^{2}-4qr^{2}\\
&-55qr+139q-32r^{3}+159r^{2}+170r-1102
\end{align*}
\begin{align*}
H_{h_1,2}={}&8p^{2}r-8p^{2}-3pq+6pr-14p+2q^{2}+32qr^{2}\\
&-56qr-154q-48r^{3}+14r^{2}+411r+620
\end{align*}
\begin{align*}
H_{h_1,3}={}&-4p^{2}r+4p^{2}+12pq-30pr+157p-8q^{2}-46qr\\
&+111q-4r^{3}+95r^{2}-65r-1188
\end{align*}
\begin{align*}
H_{h_1,4}={}&8p^{2}r-8p^{2}+12pq-24pr+111p-8q^{2}-45qr\\
&+110q+8r^{3}+100r^{2}-112r-1118
\end{align*}
\begin{align*}
H_{h_1,6}={}&3pq-6pr+34p-2q^{2}-12qr+22q-2r^{2}\\
&+11r-110
\end{align*}
\begin{align*}
H_{h_1,7}={}&-4p^{2}r+4p^{2}-12pr+87p+8qr^{2}-13qr-10q\\
&-4r^{3}+30r^{2}+22r-438
\end{align*}
\begin{align*}
H_{h_1,8}={}&9pq-18pr+77p-6q^{2}-33qr+88q+106r^{2}\\
&-111r-1016
\end{align*}
\begin{align*}
H_{h_1,9}={}&-12pq+24pr-116p+8q^{2}+44qr-104q-96r^{2}\\
&+84r+1144
\end{align*}
\begin{align*}
H_{h_1,10}={}&-16p^{2}r+16p^{2}+6pq-12pr+58p-4q^{2}-30qr\\
&+60q-8r^{3}-18r^{2}+32r+196
\end{align*}
\begin{align*}
H_{h_1,11}={}&8p^{2}r-8p^{2}-6pr+41p+4qr^{2}-10qr-9q\\
&-40r^{3}+77r^{2}+373r-420
\end{align*}
\begin{align*}
H_{h_1,12}={}&12p^{2}r-12p^{2}+6pr-41p+2qr-7q+12r^{3}\\
&-3r^{2}-31r+52
\end{align*}
\begin{align*}
H_{h_1,16}={}&4p^{2}r-4p^{2}-6pq+12pr-58p+4q^{2}+26qr\\
&-56q-4r^{3}-8r^{2}+14r+180
\end{align*}
\begin{align*}
H_{h_1,17}={}&3pq-6pr+29p-2q^{2}-13qr+28q+6r^{2}\\
&-5r-92
\end{align*}
\begin{align*}
H_{h_1,20}={}&8p^{2}r-8p^{2}+12pq-24pr+106p-8q^{2}-50qr\\
&+120q+8r^{3}+76r^{2}-100r-712
\end{align*}
\begin{align*}
H_{h_1,23}={}&-6pq+29p+4q^{2}+12qr^{2}+qr-58q+24r^{3}\\
&-102r^{2}-192r+654
\end{align*}
\begin{align*}
H_{h_1,24}={}&12pq-24pr+106p-8q^{2}-54qr+124q+48r^{2}\\
&-60r-332
\end{align*}
\begin{align*}
H_{h_1,36}={}&28r^{2}-50r-126
\end{align*}
\begin{align*}
H_{h_1,38}={}&-12p^{2}r+12p^{2}-20p-32qr^{2}+64qr+136q+44r^{3}\\
&-12r^{2}-432r-388
\end{align*}
\begin{align*}
H_{h_1,40}={}&-4qr+4q-8r^{2}-18r+262
\end{align*}
\begin{align*}
H_{h_1,45}={}&-4p^{2}r+4p^{2}-4qr^{2}+12qr+12q+44r^{3}-52r^{2}\\
&-392r-4
\end{align*}
\begin{align*}
H_{h_1,47}={}&-4qr+4q+4r^{2}-14r-10
\end{align*}
\begin{align*}
H_{h_1,65}={}&-4qr+4q-32r^{2}+28r+388
\end{align*}
\subsection{The identity for $h_2$}
Here $N_{h_2}=512$. The nonzero multipliers are as follows.
\begin{align*}
H_{h_2,0}={}&36pr^{2}-72pr+153p+18qr+63q-72r^{3}+69r^{2}\\
&+33r-174
\end{align*}
\begin{align*}
H_{h_2,1}={}&-180pq-12pr^{2}+264pr-1014p+180q^{2}-36qr^{2}+45qr\\
&-705q+288r^{3}-1875r^{2}+873r+7764
\end{align*}
\begin{align*}
H_{h_2,2}={}&16p^{3}+48p^{2}-16pq^{2}+96pqr+4pq-60pr^{2}+168pr\\
&-16p-32q^{2}r+12q^{2}-48qr^{2}+380qr+990q+344r^{3}\\
&-498r^{2}-4220r-4560
\end{align*}
\begin{align*}
H_{h_2,3}={}&-144pq-12pr^{2}+216pr-1125p+144q^{2}+18qr-591q\\
&+24r^{3}-825r^{2}+2451r+8142
\end{align*}
\begin{align*}
H_{h_2,4}={}&-144pq+24pr^{2}+144pr-909p+144q^{2}+27qr-600q\\
&-48r^{3}-774r^{2}+2628r+7698
\end{align*}
\begin{align*}
H_{h_2,6}={}&-36pq+48pr-312p+36q^{2}-42q-30r^{2}+54r\\
&+918
\end{align*}
\begin{align*}
H_{h_2,7}={}&4p^{3}+12p^{2}-4pq^{2}+24pqr-8pq-24pr^{2}+72pr\\
&-325p-8q^{2}r+12q^{2}-12qr^{2}+131qr+108q+104r^{3}\\
&-672r^{2}+754r+3054
\end{align*}
\begin{align*}
H_{h_2,8}={}&-108pq+144pr-621p+108q^{2}+27qr-558q-780r^{2}\\
&+2610r+6804
\end{align*}
\begin{align*}
H_{h_2,9}={}&144pq-192pr+996p-144q^{2}-36qr+528q+792r^{2}\\
&-2520r-7992
\end{align*}
\begin{align*}
H_{h_2,10}={}&-72pq-36pr^{2}+168pr-534p+72q^{2}-54qr-192q\\
&+72r^{3}+114r^{2}-450r-1368
\end{align*}
\begin{align*}
H_{h_2,11}={}&4p^{3}+12p^{2}-4pq^{2}+24pqr-8pq+12pr^{2}-109p\\
&-8q^{2}r+12q^{2}-48qr^{2}+158qr+117q+296r^{3}-1239r^{2}\\
&-2123r+2814
\end{align*}
\begin{align*}
H_{h_2,12}={}&36pr^{2}-72pr+153p+18qr+63q-72r^{3}+69r^{2}\\
&+33r-174
\end{align*}
\begin{align*}
H_{h_2,16}={}&72pq-96pr+498p-72q^{2}+18qr+228q+96r^{2}\\
&-396r-1296
\end{align*}
\begin{align*}
H_{h_2,17}={}&-36pq+48pr-249p+36q^{2}-9qr-114q-48r^{2}\\
&+198r+648
\end{align*}
\begin{align*}
H_{h_2,20}={}&-144pq+24pr^{2}+144pr-846p+144q^{2}-18qr-636q\\
&-48r^{3}-528r^{2}+1872r+4728
\end{align*}
\begin{align*}
H_{h_2,23}={}&4p^{3}+12p^{2}-4pq^{2}+24pqr+64pq-24pr^{2}-24pr\\
&+173p-8q^{2}r-60q^{2}+24qr^{2}+41qr+408q-160r^{3}\\
&+1116r^{2}+382r-4782
\end{align*}
\begin{align*}
H_{h_2,24}={}&-144pq+192pr-870p+144q^{2}-54qr-600q-300r^{2}\\
&+1008r+2052
\end{align*}
\begin{align*}
H_{h_2,36}={}&-24r^{2}+414r+630
\end{align*}
\begin{align*}
H_{h_2,38}={}&-16p^{3}-48p^{2}+16pq^{2}-96pqr+32pq+48pr^{2}-192pr\\
&+316p+32q^{2}r-48q^{2}+48qr^{2}-416qr-912q-320r^{3}\\
&+624r^{2}+4040r+2724
\end{align*}
\begin{align*}
H_{h_2,40}={}&-36qr+36q+276r^{2}-450r-2070
\end{align*}
\begin{align*}
H_{h_2,45}={}&-4p^{3}-12p^{2}+4pq^{2}-24pqr+8pq-24pr+4p\\
&+8q^{2}r-12q^{2}+48qr^{2}-140qr-216q-320r^{3}+996r^{2}\\
&+2936r+12
\end{align*}
\begin{align*}
H_{h_2,47}={}&-36qr+36q-48r^{2}+162r+210
\end{align*}
\begin{align*}
H_{h_2,65}={}&-36qr+36q+264r^{2}-900r-2700
\end{align*}

\section{Indices of the four rank minors}\label{sec:minorappendix}
Use the variable order $(Y_0,\ldots,Y_{11})=(a_1,\ldots,a_4,b_1,\ldots,b_4,c_1,\ldots,c_4)$ and the cubic $F=\sum_{i,j,k\text{ distinct}}a_i b_j c_k$. For a covector $\ell$, eliminate the variable with its first nonzero coefficient. For each pair $(i,j)$ form the polynomial $Y_j\partial F/\partial Y_i$ after that substitution. The coefficient matrix has column number $12i+j$. A row label $(a,b,c)$ means the coefficient of $Y_aY_bY_c$, with repetitions allowed; the coefficient is taken in the ordinary commutative monomial basis, without a factorial normalization.
For each case below, take the rows and the columns in the listed order. These rules and lists determine every entry of each minor. Indices start at zero. Together with the kernel bases in Section~\ref{sec:fourkernels}, the four determinants give the kernel dimensions. In archive~B the four cases are called \texttt{coordinate\_e1}, \texttt{difference\_w}, \texttt{complement\_sum\_r} and \texttt{generic\_difference\_line}.
\subsection{The covector $a_1$}
The covector is $(1,0,0,0,0,0,0,0,0,0,0,0)$. The selected minor has size $128$ and determinant
\[-140737488355328.\]
Its ordered column list is
\begin{quote}\small\ttfamily
1, 2, 3, 4, 5, 6, 7, 8, 9, 10, 11, 13, 14, 15, 16, 17,
18, 19, 20, 21, 22, 23, 25, 26, 27, 28, 29, 30, 31, 32, 33, 34,
35, 37, 38, 39, 40, 41, 42, 43, 44, 45, 46, 47, 49, 50, 51, 52,
53, 54, 55, 56, 57, 58, 59, 61, 62, 63, 64, 65, 66, 67, 68, 69,
70, 71, 73, 74, 75, 76, 77, 78, 79, 80, 81, 82, 83, 85, 86, 87,
88, 89, 90, 92, 93, 94, 95, 97, 98, 99, 100, 101, 102, 103, 104, 105,
106, 107, 109, 110, 111, 112, 113, 114, 115, 116, 117, 118, 119, 121, 122, 123,
124, 125, 126, 127, 128, 129, 130, 131, 133, 134, 135, 136, 137, 138, 139, 141.
\end{quote}
The ordered row list follows; read each line from left to right, then continue to the next line.
\begin{longtable}{llllll}
\toprule\endhead
$(1,1,4)$ & $(1,1,6)$ & $(1,1,7)$ & $(1,1,8)$ & $(1,1,10)$ & $(1,1,11)$\\
$(1,2,4)$ & $(1,2,6)$ & $(1,2,7)$ & $(1,2,8)$ & $(1,2,10)$ & $(1,2,11)$\\
$(1,3,4)$ & $(1,3,6)$ & $(1,3,7)$ & $(1,3,8)$ & $(1,3,10)$ & $(1,3,11)$\\
$(1,4,4)$ & $(1,4,5)$ & $(1,4,6)$ & $(1,4,7)$ & $(1,4,8)$ & $(1,4,9)$\\
$(1,4,10)$ & $(1,4,11)$ & $(1,5,6)$ & $(1,5,7)$ & $(1,5,8)$ & $(1,5,10)$\\
$(1,5,11)$ & $(1,6,6)$ & $(1,6,7)$ & $(1,6,8)$ & $(1,6,9)$ & $(1,6,10)$\\
$(1,6,11)$ & $(1,7,7)$ & $(1,7,8)$ & $(1,7,9)$ & $(1,7,10)$ & $(1,7,11)$\\
$(1,8,8)$ & $(1,8,9)$ & $(1,8,10)$ & $(1,8,11)$ & $(1,9,10)$ & $(1,9,11)$\\
$(1,10,10)$ & $(1,10,11)$ & $(1,11,11)$ & $(2,2,4)$ & $(2,2,8)$ & $(2,3,4)$\\
$(2,3,5)$ & $(2,3,8)$ & $(2,3,9)$ & $(2,4,4)$ & $(2,4,5)$ & $(2,4,6)$\\
$(2,4,7)$ & $(2,4,8)$ & $(2,4,9)$ & $(2,4,10)$ & $(2,4,11)$ & $(2,5,7)$\\
$(2,5,8)$ & $(2,5,9)$ & $(2,5,10)$ & $(2,5,11)$ & $(2,6,8)$ & $(2,6,9)$\\
$(2,6,11)$ & $(2,7,8)$ & $(2,7,9)$ & $(2,7,11)$ & $(2,8,8)$ & $(2,8,9)$\\
$(2,8,10)$ & $(2,8,11)$ & $(2,9,11)$ & $(3,4,4)$ & $(3,4,7)$ & $(3,4,8)$\\
$(3,4,9)$ & $(3,4,10)$ & $(3,4,11)$ & $(3,5,8)$ & $(3,5,9)$ & $(3,5,10)$\\
$(3,5,11)$ & $(3,6,8)$ & $(3,6,10)$ & $(3,7,8)$ & $(3,8,8)$ & $(3,8,11)$\\
$(4,4,9)$ & $(4,4,10)$ & $(4,4,11)$ & $(4,5,9)$ & $(4,5,10)$ & $(4,5,11)$\\
$(4,6,9)$ & $(4,6,10)$ & $(4,6,11)$ & $(4,7,9)$ & $(4,7,10)$ & $(4,7,11)$\\
$(4,8,9)$ & $(4,8,10)$ & $(4,8,11)$ & $(4,9,9)$ & $(4,9,10)$ & $(4,9,11)$\\
$(4,10,10)$ & $(4,10,11)$ & $(4,11,11)$ & $(5,5,10)$ & $(5,6,8)$ & $(5,6,10)$\\
$(5,7,10)$ & $(5,8,10)$ & $(5,8,11)$ & $(5,9,10)$ & $(5,10,10)$ & $(5,10,11)$\\
$(6,8,9)$ & $(6,8,11)$ &  &  &  & \\
\bottomrule\end{longtable}
\subsection{The covector $a_1-a_2$}
The covector is $(1,-1,0,0,0,0,0,0,0,0,0,0)$. The selected minor has size $128$ and determinant
\[18446744073709551616.\]
Its ordered column list is
\begin{quote}\small\ttfamily
0, 2, 3, 4, 5, 6, 7, 8, 9, 10, 11, 12, 14, 15, 16, 17,
18, 19, 20, 21, 22, 23, 24, 26, 27, 28, 29, 30, 31, 32, 33, 34,
35, 36, 38, 39, 40, 41, 42, 43, 44, 45, 46, 47, 48, 50, 51, 52,
53, 54, 55, 56, 57, 58, 59, 60, 62, 63, 64, 65, 66, 67, 68, 69,
70, 71, 72, 74, 75, 76, 77, 78, 79, 80, 81, 82, 83, 84, 86, 87,
88, 89, 90, 92, 93, 94, 95, 96, 98, 99, 100, 101, 102, 103, 104, 105,
106, 107, 108, 110, 111, 112, 113, 114, 115, 116, 118, 120, 122, 123, 124, 125,
126, 127, 128, 129, 130, 131, 132, 134, 135, 136, 137, 138, 139, 140, 141, 142.
\end{quote}
The ordered row list follows; read each line from left to right, then continue to the next line.
\begin{longtable}{llllll}
\toprule\endhead
$(1,1,4)$ & $(1,1,6)$ & $(1,1,7)$ & $(1,1,8)$ & $(1,1,10)$ & $(1,1,11)$\\
$(1,2,4)$ & $(1,2,5)$ & $(1,2,6)$ & $(1,2,7)$ & $(1,2,8)$ & $(1,2,9)$\\
$(1,2,10)$ & $(1,2,11)$ & $(1,3,4)$ & $(1,3,6)$ & $(1,3,7)$ & $(1,3,8)$\\
$(1,3,10)$ & $(1,3,11)$ & $(1,4,4)$ & $(1,4,5)$ & $(1,4,6)$ & $(1,4,7)$\\
$(1,4,8)$ & $(1,4,9)$ & $(1,4,10)$ & $(1,4,11)$ & $(1,5,6)$ & $(1,5,7)$\\
$(1,5,8)$ & $(1,5,9)$ & $(1,5,10)$ & $(1,5,11)$ & $(1,6,6)$ & $(1,6,7)$\\
$(1,6,8)$ & $(1,6,9)$ & $(1,6,10)$ & $(1,6,11)$ & $(1,7,7)$ & $(1,7,8)$\\
$(1,7,9)$ & $(1,7,10)$ & $(1,7,11)$ & $(1,8,8)$ & $(1,8,9)$ & $(1,8,10)$\\
$(1,8,11)$ & $(1,9,10)$ & $(1,9,11)$ & $(1,10,10)$ & $(1,10,11)$ & $(1,11,11)$\\
$(2,2,4)$ & $(2,2,8)$ & $(2,3,4)$ & $(2,3,8)$ & $(2,4,4)$ & $(2,4,5)$\\
$(2,4,6)$ & $(2,4,7)$ & $(2,4,8)$ & $(2,4,9)$ & $(2,4,10)$ & $(2,4,11)$\\
$(2,5,5)$ & $(2,5,7)$ & $(2,5,8)$ & $(2,5,9)$ & $(2,5,10)$ & $(2,5,11)$\\
$(2,6,7)$ & $(2,6,8)$ & $(2,6,9)$ & $(2,6,11)$ & $(2,7,8)$ & $(2,7,9)$\\
$(2,7,10)$ & $(2,7,11)$ & $(2,8,8)$ & $(2,8,9)$ & $(2,8,10)$ & $(2,8,11)$\\
$(2,9,9)$ & $(2,9,11)$ & $(2,10,11)$ & $(3,4,8)$ & $(3,4,9)$ & $(3,4,10)$\\
$(3,4,11)$ & $(3,5,10)$ & $(3,5,11)$ & $(3,6,10)$ & $(3,6,11)$ & $(3,7,8)$\\
$(4,4,9)$ & $(4,4,10)$ & $(4,4,11)$ & $(4,5,9)$ & $(4,5,10)$ & $(4,5,11)$\\
$(4,6,9)$ & $(4,6,10)$ & $(4,6,11)$ & $(4,7,9)$ & $(4,7,10)$ & $(4,7,11)$\\
$(4,8,9)$ & $(4,8,10)$ & $(4,8,11)$ & $(4,9,9)$ & $(4,9,10)$ & $(4,9,11)$\\
$(4,10,10)$ & $(4,10,11)$ & $(4,11,11)$ & $(5,5,10)$ & $(5,6,8)$ & $(5,6,10)$\\
$(5,7,10)$ & $(5,8,10)$ & $(5,8,11)$ & $(5,9,10)$ & $(5,10,10)$ & $(5,10,11)$\\
$(6,8,9)$ & $(6,8,11)$ &  &  &  & \\
\bottomrule\end{longtable}
\subsection{The covector $a_3+a_4$}
The covector is $(0,0,1,1,0,0,0,0,0,0,0,0)$. The selected minor has size $128$ and determinant
\[-9007199254740992.\]
Its ordered column list is
\begin{quote}\small\ttfamily
0, 1, 2, 4, 5, 6, 7, 8, 9, 10, 11, 12, 13, 14, 16, 17,
18, 19, 20, 21, 22, 23, 24, 25, 26, 28, 29, 30, 31, 32, 33, 34,
35, 36, 37, 38, 40, 41, 42, 43, 44, 45, 46, 47, 48, 49, 50, 52,
53, 54, 55, 56, 57, 58, 59, 60, 61, 62, 64, 65, 66, 67, 68, 69,
70, 71, 72, 73, 74, 76, 77, 78, 79, 80, 81, 82, 83, 84, 85, 86,
88, 89, 90, 92, 93, 94, 95, 96, 97, 98, 100, 101, 102, 103, 104, 105,
106, 107, 108, 109, 110, 112, 113, 114, 115, 116, 118, 120, 121, 122, 124, 125,
126, 127, 128, 129, 130, 131, 132, 133, 134, 136, 137, 138, 139, 140, 141, 142.
\end{quote}
The ordered row list follows; read each line from left to right, then continue to the next line.
\begin{longtable}{llllll}
\toprule\endhead
$(0,0,5)$ & $(0,0,6)$ & $(0,0,7)$ & $(0,0,9)$ & $(0,0,10)$ & $(0,0,11)$\\
$(0,1,5)$ & $(0,1,6)$ & $(0,1,7)$ & $(0,1,9)$ & $(0,1,10)$ & $(0,1,11)$\\
$(0,3,5)$ & $(0,3,6)$ & $(0,3,7)$ & $(0,3,9)$ & $(0,3,10)$ & $(0,3,11)$\\
$(0,4,5)$ & $(0,4,6)$ & $(0,4,7)$ & $(0,4,9)$ & $(0,4,10)$ & $(0,4,11)$\\
$(0,5,5)$ & $(0,5,6)$ & $(0,5,7)$ & $(0,5,8)$ & $(0,5,9)$ & $(0,5,10)$\\
$(0,5,11)$ & $(0,6,6)$ & $(0,6,7)$ & $(0,6,8)$ & $(0,6,9)$ & $(0,6,10)$\\
$(0,6,11)$ & $(0,7,7)$ & $(0,7,8)$ & $(0,7,9)$ & $(0,7,10)$ & $(0,7,11)$\\
$(0,8,9)$ & $(0,8,10)$ & $(0,8,11)$ & $(0,9,9)$ & $(0,9,10)$ & $(0,9,11)$\\
$(0,10,10)$ & $(0,10,11)$ & $(0,11,11)$ & $(1,1,6)$ & $(1,1,10)$ & $(1,3,6)$\\
$(1,3,7)$ & $(1,3,10)$ & $(1,3,11)$ & $(1,4,6)$ & $(1,4,7)$ & $(1,4,8)$\\
$(1,4,9)$ & $(1,4,10)$ & $(1,4,11)$ & $(1,5,6)$ & $(1,5,8)$ & $(1,5,10)$\\
$(1,5,11)$ & $(1,6,6)$ & $(1,6,7)$ & $(1,6,8)$ & $(1,6,9)$ & $(1,6,10)$\\
$(1,6,11)$ & $(1,7,8)$ & $(1,7,10)$ & $(1,7,11)$ & $(1,8,10)$ & $(1,8,11)$\\
$(1,9,10)$ & $(1,10,10)$ & $(1,10,11)$ & $(3,4,6)$ & $(3,4,7)$ & $(3,4,8)$\\
$(3,4,9)$ & $(3,4,10)$ & $(3,4,11)$ & $(3,5,10)$ & $(3,5,11)$ & $(3,6,8)$\\
$(3,6,10)$ & $(3,6,11)$ & $(3,7,8)$ & $(3,7,10)$ & $(3,8,10)$ & $(3,8,11)$\\
$(4,4,9)$ & $(4,4,10)$ & $(4,4,11)$ & $(4,5,9)$ & $(4,5,10)$ & $(4,5,11)$\\
$(4,6,9)$ & $(4,6,10)$ & $(4,6,11)$ & $(4,7,9)$ & $(4,7,10)$ & $(4,7,11)$\\
$(4,8,9)$ & $(4,8,10)$ & $(4,8,11)$ & $(4,9,9)$ & $(4,9,10)$ & $(4,9,11)$\\
$(4,10,10)$ & $(4,10,11)$ & $(4,11,11)$ & $(5,5,10)$ & $(5,6,8)$ & $(5,6,10)$\\
$(5,7,10)$ & $(5,8,10)$ & $(5,8,11)$ & $(5,9,10)$ & $(5,10,10)$ & $(5,10,11)$\\
$(6,8,9)$ & $(6,8,11)$ &  &  &  & \\
\bottomrule\end{longtable}
\subsection{The covector $a_1-a_2+2(a_3+a_4)$}
The covector is $(1,-1,2,2,0,0,0,0,0,0,0,0)$. The selected minor has size $129$ and determinant
\[146884487083177444726800384.\]
Its ordered column list is
\begin{quote}\small\ttfamily
0, 1, 2, 4, 5, 6, 7, 8, 9, 10, 11, 12, 13, 14, 16, 17,
18, 19, 20, 21, 22, 23, 24, 25, 26, 28, 29, 30, 31, 32, 33, 34,
35, 36, 37, 38, 40, 41, 42, 43, 44, 45, 46, 47, 48, 49, 50, 52,
53, 54, 55, 56, 57, 58, 59, 60, 61, 62, 64, 65, 66, 67, 68, 69,
70, 71, 72, 73, 74, 76, 77, 78, 79, 80, 81, 82, 83, 84, 85, 86,
88, 89, 90, 92, 93, 94, 95, 96, 97, 98, 100, 101, 102, 103, 104, 105,
106, 107, 108, 109, 110, 112, 113, 114, 115, 116, 117, 118, 120, 121, 122, 124,
125, 126, 127, 128, 129, 130, 131, 132, 133, 134, 136, 137, 138, 139, 140, 141,
142.
\end{quote}
The ordered row list follows; read each line from left to right, then continue to the next line.
\begin{longtable}{llllll}
\toprule\endhead
$(1,1,4)$ & $(1,1,6)$ & $(1,1,7)$ & $(1,1,8)$ & $(1,1,10)$ & $(1,1,11)$\\
$(1,2,4)$ & $(1,2,5)$ & $(1,2,6)$ & $(1,2,7)$ & $(1,2,8)$ & $(1,2,9)$\\
$(1,2,10)$ & $(1,2,11)$ & $(1,3,4)$ & $(1,3,6)$ & $(1,3,7)$ & $(1,3,8)$\\
$(1,3,10)$ & $(1,3,11)$ & $(1,4,4)$ & $(1,4,5)$ & $(1,4,6)$ & $(1,4,7)$\\
$(1,4,8)$ & $(1,4,9)$ & $(1,4,10)$ & $(1,4,11)$ & $(1,5,6)$ & $(1,5,7)$\\
$(1,5,8)$ & $(1,5,9)$ & $(1,5,10)$ & $(1,5,11)$ & $(1,6,6)$ & $(1,6,7)$\\
$(1,6,8)$ & $(1,6,9)$ & $(1,6,10)$ & $(1,6,11)$ & $(1,7,7)$ & $(1,7,8)$\\
$(1,7,9)$ & $(1,7,10)$ & $(1,7,11)$ & $(1,8,8)$ & $(1,8,9)$ & $(1,8,10)$\\
$(1,8,11)$ & $(1,9,10)$ & $(1,9,11)$ & $(1,10,10)$ & $(1,10,11)$ & $(1,11,11)$\\
$(2,2,4)$ & $(2,2,8)$ & $(2,3,4)$ & $(2,3,8)$ & $(2,4,4)$ & $(2,4,5)$\\
$(2,4,6)$ & $(2,4,7)$ & $(2,4,8)$ & $(2,4,9)$ & $(2,4,10)$ & $(2,4,11)$\\
$(2,5,5)$ & $(2,5,7)$ & $(2,5,8)$ & $(2,5,9)$ & $(2,5,10)$ & $(2,5,11)$\\
$(2,6,7)$ & $(2,6,8)$ & $(2,6,9)$ & $(2,6,10)$ & $(2,6,11)$ & $(2,7,8)$\\
$(2,7,9)$ & $(2,7,10)$ & $(2,7,11)$ & $(2,8,8)$ & $(2,8,9)$ & $(2,8,10)$\\
$(2,8,11)$ & $(2,9,9)$ & $(2,9,11)$ & $(2,10,11)$ & $(3,4,8)$ & $(3,4,9)$\\
$(3,4,10)$ & $(3,4,11)$ & $(3,5,10)$ & $(3,5,11)$ & $(3,6,10)$ & $(3,6,11)$\\
$(3,7,8)$ & $(4,4,9)$ & $(4,4,10)$ & $(4,4,11)$ & $(4,5,9)$ & $(4,5,10)$\\
$(4,5,11)$ & $(4,6,9)$ & $(4,6,10)$ & $(4,6,11)$ & $(4,7,9)$ & $(4,7,10)$\\
$(4,7,11)$ & $(4,8,9)$ & $(4,8,10)$ & $(4,8,11)$ & $(4,9,9)$ & $(4,9,10)$\\
$(4,9,11)$ & $(4,10,10)$ & $(4,10,11)$ & $(4,11,11)$ & $(5,5,10)$ & $(5,6,8)$\\
$(5,6,10)$ & $(5,7,10)$ & $(5,8,10)$ & $(5,8,11)$ & $(5,9,10)$ & $(5,10,10)$\\
$(5,10,11)$ & $(6,8,9)$ & $(6,8,11)$ &  &  & \\
\bottomrule\end{longtable}

\end{document}